\documentclass[manuscript, nonacm]{acmart}
\AtBeginDocument{%
  }

\usepackage{xcolor} % in preamble
\usepackage{multirow}
\usepackage{tabularx}

\renewcommand\footnotetextcopyrightpermission[1]{}
\setcopyright{none}
\begin{document}

\title{Tracing Prompt-Level Trajectories to Understand Student Learning with AI in Programming Education}

\author{Tianyu Shao}
\email{shao122@purdue.edu}
\affiliation{%
  \institution{Purdue University}
  \country{USA}
}

\author{Miguel Feijóo-García}
\email{Mfeijoog@purdue.edu}
\affiliation{%
  \institution{Purdue University}
  \country{USA}
}

\author{Yi Zhang}
\email{zhan3050@purdue.edu}
\affiliation{%
  \institution{Purdue University}
  \country{USA}
}

\author{Hugo Castellanos}
\email{hcastel@purdue.edu}
\affiliation{%
  \institution{Purdue University}
  \country{USA}
}

\author{Tawfiq Salem}
\email{tsalem@purdue.edu}
\affiliation{%
  \institution{Purdue University}
  \country{USA}
}

\author{Alejandra Magana}
\email{admagana@purdue.edu}
\affiliation{%
  \institution{Purdue University}
  \country{USA}
}

\author{Tianyi Li}
\email{li4251@purdue.edu}
\affiliation{%
  \institution{Purdue University}
  \country{USA}
}

\renewcommand{\shortauthors}{Shao et al.}

% Define the note command
\newcommand{\note}[1]{\textcolor{teal}{\textit{[Note: #1]}}}

\newcommand{\tianyi}[1]{\textcolor{teal}{\textit{[TL: #1]}}}
\newcommand{\carol}[1]{\textcolor{magenta}{\textit{[TS: #1]}}}

\begin{abstract}
% This paper presents an empirical study of how undergraduate students use publicly available, general-purpose large language models (LLMs) such as ChatGPT in timed problem-solving from start to end. %As LLMs become increasingly accessible, the computing education community has raised concerns and opportunities around their use, with the shared goal of enhancing student learning. A key step toward this goal is understanding how students naturally incorporate AI into their problem-solving processes when no instructor regulations are imposed. Such insights highlight where students need more support, reveal their mental models for completing assignments, and identify leverage points for meaningful scaffolding of AI-assisted learning. 
% We analyzed chat histories and assignment submissions from 163 students in an introductory Python course, where students used the free tier of ChatGPT to implement a number-guessing game with Python. Our analysis of prompt-level strategies, interaction trajectories, and code comparisons shows students' AI use behaviors ranging from full delegation to iterative refinement. We contrast these patterns with assignment outcomes and draw design implications for educational AI systems that foster personalized, productive student–AI collaboration.

  As AI tools such as ChatGPT enter programming classrooms, students encounter differing rules across courses and instructors, which shape how they use AI and leave them with unequal capabilities for leveraging it. We investigate how students engaged with AI in an introductory Python assignment, analyzing student–LLM chat histories and final code submissions from 163 students. We examined prompt-level strategies, traced trajectories of interaction, and compared AI-generated code with student submissions. We identified trajectories ranging from full delegation to iterative refinement, with hybrid forms in between. Although most students directly copied AI-generated code in their submission, many students scaffolded the code generation through iterative refinement. We also contrasted interaction patterns with assignment outcomes and course performance. Our findings show that prompting trajectories serve as promising windows into students’ self-regulation and learning orientation. We draw design implications for educational AI systems that promote personalized and productive student-AI collaborative learning.
  
\end{abstract}



%   \caption{\note{To be replaced by a figure that captures the key findings}}
%   \Description{tbd}


\maketitle

\section{Introduction}
The rapid rise of Large Language Models (LLMs) such as ChatGPT, Gemini, and Claude has reshaped programming education by making on-demand code generation, conceptual explanations, and debugging support widely accessible~\cite{agbo2025computing, garcia2025teaching, vadaparty2024cs1}. This availability offers significant benefits, but also poses new challenges for educators: understanding how students actually learn with LLMs, identifying their evolving needs, and adapting instruction accordingly. At the same time, industry increasingly expects graduates to use LLMs effectively in their work, creating pressure for curricula to integrate these tools. Yet concerns persist that instant access to solutions may bypass the cognitive struggle and problem-solving processes essential for mastering programming. As a result, many undergraduate courses continue to discourage or restrict LLM use, and course-level policies vary widely.

Studies have shown that while LLMs lower barriers to code generation and debugging, they also risk undermining essential learning processes. Novice programmers often struggle to craft effective prompts, which exposes fragile programming skills and limited prompt literacy~\cite{nguyen2024how, kumar2024guiding}. In some cases, students have passed exams with ChatGPT’s help despite weak preparation, a pattern that suggests reliance on AI can replace foundational practice rather than complement it~\cite{shoufan2023can}. Reliance on LLMs for debugging has also been shown to reduce opportunities for developing independent problem-solving skills~\cite{pirzado2024navigating}, while hallucinations and partial accuracy cast doubt on the reliability of AI support in STEM domains~\cite{ding2023students, arantes2024understanding}. These findings point to a central tension: LLMs provide powerful cognitive scaffolding, yet they can displace the productive struggle that is critical to mastering programming.

Recent research has begun to examine this emerging landscape from multiple perspectives. One line of work explores how students use LLMs in programming courses, analyzing prompt types, usage patterns, and motivations for turning to AI~\cite{oliveira2025can, kazemitabaar2024how, amoozadeh2024student}. These studies reveal that students seek a range of support, from complete solutions to targeted fixes or explanations, but they largely capture single points of interaction rather than full sessions. In practice, prompting behaviors shift dynamically within a task, yet few studies trace how these behaviors evolve across the entire problem-solving process. Another thread investigates how to help students engage more effectively with LLMs, for example, by scaffolding prompt formulation or encouraging reflection on AI outputs rather than treating them as black-box solutions~\cite{kumar2024guiding, nguyen2024how}. These interventions highlight the value of prompt literacy and metacognitive awareness but usually emphasize local improvements rather than holistic interaction trajectories. A third line of work designs collaborative interfaces that assign LLMs roles such as feedback providers, brainstorming partners, or debugging assistants~\cite{cope2024platformed, frazier2024customizing, vadaparty2024cs1}. While these efforts illustrate the potential of AI as an interactive learning companion, they often assume fixed roles, whereas actual student use blends delegation, repair, and reflection within a single task.

Nonetheless, we still know little about how students actually use LLMs to navigate an entire problem-solving episode under authentic constraints. Most existing studies either isolate fragments of interaction or spread usage over extended timeframes, leaving unexamined the moment-to-moment strategies students deploy when solving a complete task from start to finish. Yet this end-to-end perspective is essential. Timed, single-session activities simulate the realities of student practice, where work is often completed under time pressure and efficiency takes precedence over exploration. Such contexts reveal how students manage cognitive load, shift between delegation and independent effort, and converge—or diverge—in their final solutions. By zooming in on a single, time-bounded assignment, we can trace interaction trajectories as they unfold, offering unique insight into how LLMs are appropriated as cognitive partners when the stakes of immediacy, completion, and correctness are most salient.
We investigated student-LLM interaction trajectories within an in-class programming activity in an introductory Python course (Spring 2025) where 163 students were asked to solve a number-guessing game within a single lecture session, with the option of using an LLM for support. We collected both their chat transcripts and their final code submissions, enabling us to trace the full arc of interaction from initial prompt to completed solution. Our analysis addresses three research questions:

\textbf{RQ1:} How do students’ prompts to LLMs evolve, and what recurring interaction trajectories emerge?

\textbf{RQ2:} How closely do students’ submissions align with LLM-generated code, and to what extent do their solutions converge with one another?

\textbf{RQ3: }How do interaction trajectories relate to code similarity, and how do these patterns affect their final code submission quality?

Across 146 valid submissions and 662 prompts, we identify eight distinct interaction trajectories that reflect how students distributed cognitive effort between themselves and the LLM. We find strong convergence in student code, with over 80\% directly reusing AI-generated solutions, though meaningful variation persisted at the margins. Students who fully outsourced to the LLM often performed as well as those who engaged more strategically, highlighting tensions between efficiency and reflection. Our results reveal how LLM-supported programming increasingly resembles a form of rapid prototyping: generate drafts, test them against requirements, and iterate through cycles of adaptation or debugging.

This work makes the following contributions:

\begin{enumerate}
    \item \textbf{A typology of interaction trajectories.} From 662 prompts across 146 valid submissions, we identify eight distinct trajectories that capture how students delegated, adapted, or debugged with LLMs across a single, full problem-solving process.

    \item \textbf{A dataset of paired LLM transcripts and student submissions.} To our knowledge, this is among the first classroom datasets that capture prompt-level interactions and final code under a single and complex problem-solving process from start to end. 

    \item \textbf{Design implications for LLM-supported learning.} We reframe LLM-supported programming as cycles of prototyping, offering implications for assignment design, assessment, and equity that emphasize iteration and reflection as crucial learning goals.
\end{enumerate}

\subsection*{Positionality Statement}
This study was conducted by an interdisciplinary team with backgrounds in Human–Computer Interaction (HCI), Engineering Education, Natural Language Processing (NLP), and Computer Vision (CV). The course instructor (a CV researcher) and the teaching assistant (an NLP researcher) designed and administered the assignment and facilitated data collection. Data analysis was led by two HCI researchers, and the two engineering education students assisted with codebook applications and evaluation, with guidance from a senior engineering education researcher. Our varied disciplinary orientations provided both technical and pedagogical insights, enabling us to approach student–LLM interactions from multiple perspectives. At the same time, these backgrounds inevitably shaped the lenses through which we interpreted student practices, emphasizing problem-solving strategies, learning processes, and the role of human-AI interaction in education.
\section{Related Work}

\subsection{LLM in Programming Education}
Large Language Models (LLMs) such as ChatGPT and code-specialized assistants like OpenAI Codex are increasingly integrated into programming contexts, spanning introductory CS1 courses to advanced problem-solving and design tasks. Prior studies have examined novice programmers~\cite{nguyen2024how, kazemitabaar2023studying, vadaparty2024cs1, amoozadeh2024student}, struggling or underprepared students~\cite{shoufan2023can}, and, in some cases, more experienced learners in professional or design settings~\cite{liang2024large}. Educational deployments typically center on Python assignments, debugging activities, multiple-choice questions, fill-in-the-blank questions, or open-ended projects where LLMs generate, explain, or refine code~\cite{kazemitabaar2024how, oliveira2025can, ambikairajah2024chatgpt}.

Findings across these contexts are mixed but converge on several themes. LLMs lower entry barriers by providing on-demand explanations, examples, and personalized guidance~\cite{garcia2025teaching, penney2023assessing, lyu2024evaluating}, and they often improve assignment outcomes, particularly for novices and first-time users~\cite{lyu2024evaluating, kazemitabaar2023studying}. However, generated code frequently contains errors or misleading explanations that students struggle to detect~\cite{ali2023assessment, misanchuk2023chatgpt, pirzado2024navigating}. Over-reliance may lead to code homogeneity and reduce opportunities for debugging practice~\cite{amoozadeh2024student, budhiraja2024its, joshi2024chatgpt}. Moreover, benefits are uneven: some learners use LLMs strategically, while others copy outputs verbatim or are limited by weak prompting skills~\cite{nguyen2024how, kazemitabaar2024how}.

% Although LLMs are reshaping programming education by lowering barriers and expanding support, they also introduce risks of error propagation, over-reliance, and inequity in different contexts and overtime. Building on these insights, our study investigates how such impacts unfold over the course of an uninterrupted, end-to-end problem-solving process.

\subsubsection{Patterns of Student–LLM Interaction}
Prior research shows that novice students adopt a wide range of prompting strategies when using LLMs for learning. Some request complete solutions~\cite{kazemitabaar2024how}, others break tasks into subgoals or debugging prompts~\cite{amoozadeh2024student, oliveira2025can}, and still others seek explanations or planning support~\cite{shoufan2023can}. More advanced strategies include iterative refinement, drafting pseudocode, or embedding input–output examples in prompts~\cite{kazemitabaar2024how, haindl2024students}. Timing also varies: some students turn to the LLM immediately (``LLM-first''), while others attempt the task independently before seeking assistance (``self-first'')~\cite{kumar2024guiding, vadaparty2024cs1}.
These studies highlight important orientations but also recurring challenges. Novices often struggle to formulate effective prompts~\cite{lyu2024evaluating, nguyen2024how}, identical queries can produce inconsistent outputs~\cite{nguyen2024how, zhou2024developing}, and many students over-rely on LLMs by copying code wholesale~\cite{kazemitabaar2023studying, amoozadeh2024student}. Prompting behaviors are also dynamic rather than fixed: students frequently shift mid-task, oscillating between independent effort and full reliance on the model~\cite{kazemitabaar2024how}.

Current evidence thus provides valuable categories and examples of how novices interact with LLMs, typically in settings where they have not yet mastered the relevant concepts. In this work, we extend this literature by examining whether similar patterns appear among students who have recently learned and practiced new programming concepts. We further trace how students' interaction with LLM unfolds as trajectories within a single problem-solving task: how their prompting strategies evolve over time and how these trajectories relate to their course performance.
% Research on students' use of Large Language Models (LLMs) in programming education has expanded rapidly, spanning methodological innovations, analyses of student–AI interactions, and investigations of learning outcomes. %Prior studies provide valuable insights into how learners prompt LLMs~\cite{}, how these prompts shape the code they produce~\cite{}, and what challenges and opportunities arise from integrating AI into coursework~\cite{}. At the same time, the literature remains fragmented: some work focuses primarily on prompt types, others investigate code quality or performance outcomes. Few connect these dimensions across the full cycle from interaction to submission. In this section, we review four strands of related work that together our study: methodologies for studying student–AI interaction, patterns of student use of LLMs in programming courses, impacts of LLMs on learning outcomes, and learner variability frameworks that help explain why interaction styles differ. Each strand highlights open questions that our work addresses by linking prompt evolution, code similarity, and assignment quality within a unified analysis.  

\subsubsection{Methodologies for Studying Student–AI Interaction}
Prior research has examined student–AI interaction across varied contexts and timescales. Many controlled lab studies use simplified programming problems to test the effects of prompt types, guidance conditions, or AI access, yielding rigorous but tightly scoped findings~\cite{nguyen2024how, shoufan2023can, kazemitabaar2023studying, kumar2024guiding}. In contrast, classroom-based research offers greater ecological validity. Some studies embed LLMs directly into curricula for weeks or semesters~\cite{vadaparty2024cs1, lyu2024evaluating}, while others capture authentic use through logged interactions in assignments using plugins or custom platforms~\cite{amoozadeh2024student, kazemitabaar2024how, salminen2024using}. These approaches reveal how students use LLMs in practice but often involve interventions that steer them toward particular strategies, limiting observation of unregulated behavior.
Studies also differ in timescales: in-lab studies typically spans a single session lasting a few hours~\cite{nguyen2024how}, while classroom studies extend across weeks~\cite{liang2024large} or entire semesters~\cite{vadaparty2024cs1}. Although classroom studies provide more realistic insights, they are often confounded by the broader instructional and personal contexts in which LLM use unfolds.

In this work, we take a complementary perspective by zooming in on a standalone, single-session problem-solving process. We analyze how students used LLMs from start to finish within a timed in-class activity, without instructor-designed guidance or scaffolding on when and how to use LLM, to capture their spontaneous trajectories of interaction.

\subsection{Cognitive and HCI Perspectives on Problem-Solving Under Pressure}
Prior work has conceptualized digital tools as cognitive partners that reshape problem-solving, while HCI research highlights how time pressure transforms user strategies and outcomes. Together, these perspectives provide the foundation for analyzing student–LLM interaction trajectories.

\subsubsection{Digital Tools as Cognitive Partners}
Theories of cognitive tools for learning~\cite{kommers1992cognitive}, grounded in Vygotsky’s theory of cognitive development~\cite{vygotsky1978mind}, emphasize that digital systems can augment human thought, enabling learners to engage in higher-order reasoning tasks that might otherwise be infeasible~\cite{reinhold2024learning}. LLMs exemplify such tools, capable of generating, summarizing, and transforming information at scale~\cite{tankelevitch2024metacognitive}.

This partnership introduces the dynamics of cognitive offloading and reallocation~\cite{cabrera2023improving}. Students can strategically outsource low-level tasks—such as recalling syntax or generating boilerplate code—to the LLM, thereby conserving working memory for higher-order processes like planning, hypothesis testing, or evaluating solutions~\cite{ma2025towards}. While critics warn of over-reliance leading to shallow engagement, recent perspectives view LLM-supported problem-solving as a process of redistributing cognitive effort across human and machine in specific contexts~\cite{spitzer2025human, cabrera2023improving}.

Empirical studies of student engagement with digital tools reveal diverse strategies, from verification~\cite{ambikairajah2024chatgpt, kazemitabaar2024how} (e.g., checking algebraic solutions against a graphing tool) to exploration (e.g., visualizing relationships)~\cite{misanchuk2023chatgpt} to full workflow integration~\cite{vadaparty2024cs1}. LLMs expand this landscape further, serving as brainstorming partners~\cite{misanchuk2023chatgpt}, Socratic questioners~\cite{shoufan2023can}, collaborative writers~\cite{su2023collaborating}, or simulators~\cite{barambones2024chatgpt}. A student's trajectory of interaction with the LLM thus reflects their chosen strategy for incorporating this cognitive partner into their problem-solving process.

\subsubsection{Effects of Time Constraints on Strategy and Behavior}
Time pressure is a critical contextual factor in HCI research, shown to heighten cognitive load, narrow attentional focus, and push reasoning from systematic analysis toward heuristic shortcuts~\cite{wu2022time}. Under strict deadlines, users tend to avoid deep exploration of interfaces and instead satisfice with ``good enough'' solutions.

In the context of LLM use, time constraints shift students' goals from comprehension to completion~\cite{sellen2025effects}. Each prompt becomes a time-sensitive decision: is it quicker to tackle the subproblem independently, or to engage the LLM while risking delays in prompt refinement, response evaluation, and integration? Students must juggle the intrinsic cognitive demands of problem solving with the extraneous overhead of interaction management—articulating prompts~\cite{SWELLER201137}, verifying outputs, and adapting responses. Time limits amplify this trade-off, making inefficient interaction management directly detrimental to performance~\cite{ordonez1997decisions}. In this sense, the course activity employed in this work simulates the ``deadline-fighting'' behaviors students often display in authentic academic settings.

\section{Methods}
To investigate how students leverage LLMs to support complex problem-solving from start to end with time pressure, we conducted an in-class activity where students were tasked to complete a number-guessing game, with the option to use LLM support. We collected both their chat histories with the LLM and their final code submissions, all completed within the same session. This study was approved by the Institutional Review Board (IRB) of the authors' institution.

\subsection{Course Contexts and Participants}
% \note{number of students, student backgrounds (year, major, demographic info, etc.), course level and key topics, offered every semester/year?, prerequisites? }
The in-class activity was conducted in an introductory Python programming course at a large R1 research university during the Spring 2025 semester. The course was delivered over 16 weeks with two 50-minute lectures and one 2-hour laboratory per week. The primary topics include Variables and Data Types, Selection Statements (if/else), Loops (while, for), Lists and Strings, File I/O, and Basic GUI elements. There are no programming prerequisites, and it primarily enrolls first- and second-year students majoring in Computer and Information Technology and Cybersecurity. 

There were a total of 163 students enrolled in the course, including 124 males (76.2\%) and 39 females (23.8\%). Reported ethnicity included 32.0\% White, 28.5\% Asian, 11.3\% International, 4.7\% two or more races, 4.3\% Black or African American, 3.5\% Hispanic/Latino, and 15.6\% undisclosed. Most students majored in Cybersecurity (53.1\%) or Computer and Information Technology (21.5\%), with others in Data Analytics (6.3\%) and a variety of smaller majors. The demographic backgrounds of students in this course are aligned with the overall demographics of CIT and Cybersecurity majors at the authors' institution (approximately 21\% female, 39\% White, 24\% Asian, 6\% Black, and 9\% Latino).

\subsection{Course Activity Design} 
% \note{briefly describe the structure of the assignment. The full assignment instructions can be included in supplementary materials, so no need to go full detail here}
The course activity involves completing a programming assignment, with the option to use LLM tools for assistance. At the end, students are required to submit both their assignment solution and chat history with the LLM. 

\paragraph{Assignment Design and Rationale} The assignment was structured as an interactive Number Guessing Game implemented in Python, in which the program randomly selects a number within a specified range and the player attempts to guess it, receiving feedback (e.g., too high or too low) until the number is correctly identified or the allowed attempts are exhausted. Students were provided with a partially completed code skeleton containing function headers, placeholder comments, and structural steps, while key implementation details were intentionally omitted. Their task was to complete the missing sections to produce a fully functional game, including prompting the player for input, generating random numbers based on difficulty level, implementing a guessing loop with conditional logic, tracking attempts and outcomes, supporting multiple rounds, and displaying a game summary. This assignment reflects the course objectives of problem decomposition, algorithmic thinking, and mastery of core programming constructs in an engaging, applied context.

The assignment was intentionally designed to limit the effectiveness of directly generated LLM solutions and to encourage student engagement with the underlying problem-solving process. The instructor implemented the following design elements to support this goal. First, students were provided with a partially completed code skeleton, which required them to integrate their solutions into predefined placeholders rather than submitting any program that implements the same game logic. Second, the task combined multiple functional requirements -- including input/output, loops, conditionals, lists, and randomness -- that had to be completed separately and then integrated into a single project. Third, the assignment included features that are error-prone if implemented without careful reasoning, such as updating attempts, breaking out of loops, and tracking multiple rounds of gameplay. These aspects are points where LLM-generated solutions often contain logical inconsistencies or incomplete handling. 
Our objective was not to make the task unsolvable by LLMs, but rather to design it in a way that encourages students to engage meaningfully with the problem, practice core programming skills, and learn from the process.

\paragraph{LLM Use Protocol} %\note{instructions given, platform (free ChatGPT 3.5-turbo), any constraints.}
Students were permitted—but not required—to use the free version of ChatGPT (GPT-3.5-turbo) to support their problem-solving process. They were also informed that they could use any other LLM of their choice, with no restrictions on how they use and interact with the LLM tools.

\paragraph{Submission protocol} %\note{instructions for submitting code and chat logs}
At the end of the lecture session, students were required to submit two artifacts to the course LMS: (1) the completed Python source code (.py file) and (2) a transcript of their interaction with the LLM. For the transcript, students were instructed to copy and paste the entire conversation from their chat interface into a Word document and upload it in .doc or .docx format. In addition, they were required to provide a shareable link generated by the LLM platform to ensure that the full interaction was captured.

\subsection{Procedure}
% \note{describe the entire procedure for data collection. This should include: when and how instructions are given to students (e.g. were they informed of this class activity before hand? were they asked to bring a laptop to the class and coded in the classroom? where were instructions posted? what tools were used for coding? We will need to include the actual assignment in the supplementary materials when submitting) Add a figure. }
The data collection was conducted at Week 14 during a 50-minute lecture period. At this time, students had gained sufficient experience with the knowledge required to complete the assignment: conditionals, loops, and lists. Students were instructed to bring their personal laptops to the class.

At the beginning of the activity, the course instructor introduced the programming assignment and outlined its learning objectives. The teaching assistant (TA) then executed the completed program in a terminal to demonstrate the game’s behavior, using test cases to illustrate gameplay without exposing the underlying code. Following this, the TA demonstrated how to save and export chat histories using the free version of ChatGPT (GPT-3.5-turbo). Students were provided with the assignment description, a code template, and separate instructions for obtaining and submitting their chat transcripts, all distributed via the course Learning Management System (LMS).

At the end of the lecture session, all students were asked to submit their (1) completed Python source code (.py file) and (2) their conversation history transcript with the LLM to the LMS before leaving the class.

\paragraph{Data Collection and Anonymization}
Submissions were analyzed only after final course grades had been posted. A TA, trained under the IRB protocol and in compliance with the Family Educational Rights and Privacy Act (FERPA) manually inspected each submission. The inspection process began by verifying completeness, cross-referencing the shared LLM link against the transcript file, and confirming the presence of the Python file. Incomplete submissions (e.g., missing files) or invalid ones (e.g., transcripts containing only code) were excluded.

Valid submissions were then anonymized. The TA redacted all personally identifiable information (PII) from the Python code, LLM transcripts, and any screenshots. For text, PII was removed or replaced with ``[USER]''; for images, black boxes were applied to obscure identifying details. The same TA double-checked the anonymized dataset during the initial data-tagging phase, providing final verification before the data were shared with coauthors for analysis.

\subsection{Qualitative Analysis of Conversation Histories}
To investigate students’ prompting behaviors and how they evolved during the problem-solving process (RQ1), we conducted a qualitative coding and sequence analysis of the valid LLM conversation transcripts. In parallel, because RQ2 concerns convergence between LLM outputs and students’ final code, we also developed a separate code similarity codebook.

\begin{figure}
    \centering
    \includegraphics[width=0.9\linewidth]{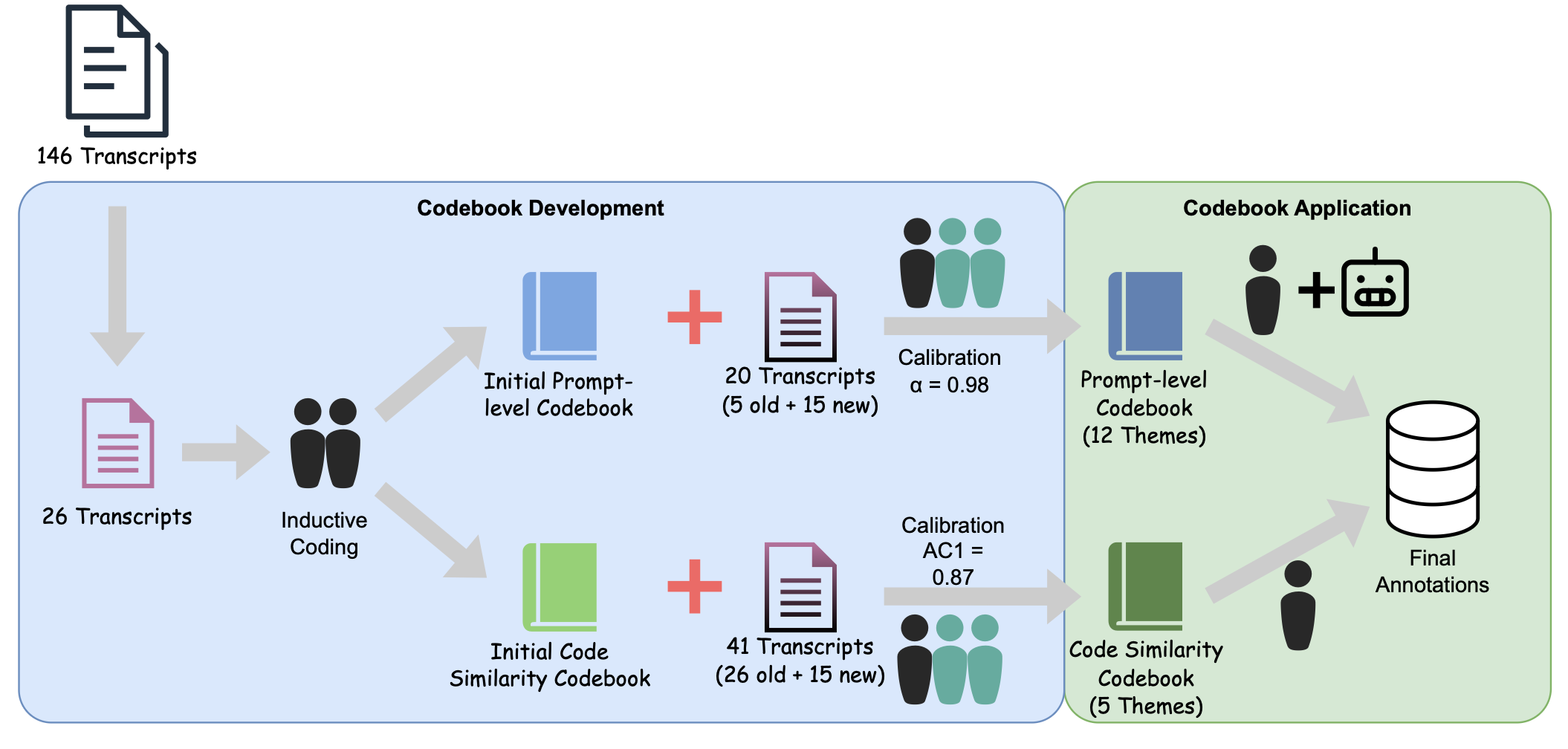}
    \caption{An illustration of the qualitative analysis process.}
    \label{fig:placeholder}
\end{figure}

\subsubsection{Codebook Development.} Two authors conducted initial inductive coding on a subset of 26 transcripts.  Each author first familiarized themselves with the dataset by reviewing each transcript a few times. They then independently performed two analysis tasks: (1) labeling individual prompts according to students' prompt traits and synthesizing recurring categories across the 26 submissions, and (2) comparing the AI-generated code in the transcripts with the students' submitted Python files to determine the extent to which the final code was derived from LLM-generated content.
For the prompt-level analysis, each student message was treated as a unit of analysis and annotated with one or more applicable prompt themes. The sequential order of prompts (Round 1, Round 2, ...) was also recorded to allow reconstruction of interaction trajectories.  
For the code-similarity analysis, each submission was assigned a single label. %because this scheme was outcome-oriented, reflecting the relationship between the AI output and the final artifact. %Unlike the prompt-level scheme, which covered multiple turns, code similarity represented the overall endpoint of the interaction.

The two authors then met to compare coding decisions, resolve disagreements through discussion, and collaboratively refine and synthesize the codes into higher-level themes for both tasks. This process produced two codebooks: one for characterizing prompts and one for assessing code similarity.
% The prompt-level codebook identified 12 themes capturing distinct prompt traits, and the code-similarity codebook identified 5 levels of similarity.

To evaluate the robustness of the codebooks, one of the authors who developed the codebook, along with two additional authors not involved in codebook construction, applied the codebooks to a set of 20 transcripts. For calibration, this includes five transcripts from the original 26 and 15 new transcripts. 
Inter-rater agreement was assessed using Krippendorff's alpha~\cite{hayes2007answering}. For the prompt-level codebook, reliability was high ($\alpha = 0.98$ across 12 themes), indicating substantial consistency~\cite{marzi2024k}. For the code-similarity codebook, Krippendorff's alpha was lower ($\alpha = 0.43$ across five themes), even though percent agreement within each theme exceeded 90\%. This discrepancy likely reflects the mutually exclusive nature of the five themes and the uneven distribution of codes across categories. To address the prevalence and bias issues inherent in chance-corrected agreement measures, we employed Gwet’s AC1~\cite{gwet2001handbook}, which yielded 0.87, indicating substantial agreement~\cite{wongpakaran2013comparison}.
This second round of coding surfaced ambiguities in the category definitions but did not yield new themes. The final versions of the codebooks are presented in Table~\ref{tab:prompt-codebook} and Table~\ref{tab:code-similarity}). 

\subsubsection{Codebook Application} Given the high level of agreement among human coders, we employed GPT-5 to annotate the remaining student samples using the finalized \textbf{prompt-level codebook} (Table~\ref{tab:prompt-codebook}) and manually verified and corrected the LLM-generated annotations. 

After iterative prompt engineering, the final prompt used to apply the codebook includes the following information: (1) detailed definitions and examples for each theme, (2) clear instructions to assign one or more labels to each student prompt, and (3) formatting constraints to standardize output for downstream analysis. Multiple trial runs were conducted on a small subset of transcripts to refine the prompt design before applying it to the full dataset. The final prompt is in appendix~\ref{app: gpt_training_prompt}. Due to the low accuracy for Themes 1-4, GPT-5 was only employed for Themes 5-12. Themes 1-4 were manually coded. 
We assessed the model's performance by calculating inter-rater agreement on a subset of 41 submissions (91 prompts) that had been independently coded by human annotators during the codebook development process. Theme-wise Krippendorff’s \(\alpha\) ranged from 0.21 to 1, while the bias-corrected Gwet’s AC1 ranged from 0.58 to 1. The overall agreement across themes was \(\alpha=0.66\) and AC1 \(=0.91\). According to Gwet’s guidelines~\cite{gwet2001handbook}, these values indicate that the model achieved substantial to almost perfect agreement with human coders. 

Based on these results, we continued to employ the same prompt and GPT-5 to annotate the remaining transcripts. To ensure validity, one of the original coders conducted a full manual inspection of all transcripts and annotations to verify and correct the LLM-generated annotations. 

For the \textbf{code-similarity codebook}, all annotations were conducted by one of the authors who developed the codebook. We initially piloted using GPT-5 to apply the code-similarity codebook; however, agreement with human annotations was too low to be useful. This is probably related to the way AI-generated code appeared in dispersed segments throughout the chat history, thus requiring holistic judgments about integration, overlap, and originality in the student's Python program. %Consequently, all submissions were manually coded.

% \subsubsection{Trajectory Reconstruction.} After finalizing the prompt-level and code-similarity codebooks, we aggregated prompt-level themes into interaction trajectories of each submission. 

% For prompt-based trajectories, we focused on Themes~5–12 from the prompt-level codebook, as these captured students' orientations toward the LLM (e.g., requesting full solutions, seeking feature-level code, or troubleshooting). Themes~1–4, which primarily reflected context provision such as pasting assignment instructions or skeleton code, were excluded from this stage. Each transcript’s sequence of themes across rounds was examined to determine whether it met the conditions for one of the predefined trajectory categories. Rules specified, for example, whether a particular theme needed to appear consistently across all rounds, whether themes had to shift from one type to another over time, or whether alternating patterns were required. 

% For hybrid trajectories, we combined the prompt-level sequences with the code-similarity outcomes. Here, the trajectory assignment depended on both the presence of specific prompt themes and the final code-similarity label assigned to the student’s submission. For instance, transcripts containing a single Theme~5 request were linked with code-similarity categories A–E to distinguish between cases where students submitted direct copies, modified examples, or original work.   

\subsection{Code Clustering Analysis}
To investigate whether LLM use led to homogeneity in student code submissions (RQ2), we applied k-means clustering to the submitted Python programs. The analysis examined convergence and divergence across the cohort using three complementary feature sets.

The first feature set was rubric-based, derived from the grading rubric in Table~\ref{tab:code-rubric}, which scored each submission on functional correctness and coding best practices. The second feature set captured structural characteristics, including surface-level measures (lines, characters, comments, blank lines) and Abstract Syntax Tree (AST)–based metrics (functions, classes, conditionals, loops, imports, variables, cyclomatic complexity). Additional patterns were identified with regular expressions (e.g., print statements, input handling, literals, random number usage, presence of a main function), yielding 26 measures in total. The third feature set used semantic embeddings from CodeBERT, a pre-trained transformer for code representation. Each script was embedded as a 768-dimensional vector, with versions both including and excluding comments.

All features were standardized to zero mean and unit variance prior to clustering. For each feature set, we determined the optimal number of clusters using both the Elbow method and Silhouette scores, selecting the \textit{k} that maximized cohesion and separation. Beyond individual analyses, we also performed combined clustering on structural and semantic features to assess whether integrated representations produced more coherent groupings.
\section{Results}
All 163 students submitted their work at the end of the lecture session. Of these, 17 submissions were excluded from analysis due to missing files or deviations from the submission instructions. The final dataset, therefore, consists of 146 submissions, each containing a Python program and a transcript of the student's chat history with the LLM.
\begin{table}[!htb]
\centering
\small
\begin{tabularx}{0.8\linewidth}{@{}c *{9}{>{\centering\arraybackslash}X}@{}}
\toprule
\textbf{Rounds}     & 1  & 2  & 3  & 4  & 5  & 6  & 7  & 8  & 9  \\
\midrule
\textbf{\# Students} & 59 & 40 & 20 & 12 & 9  & 0  & 2  & 3  & 1  \\
\bottomrule
\end{tabularx}
\caption{Number of students by rounds of conversation with the LLM}
\label{tab:rounds-distribution}
\end{table}
\subsection{Student-AI Interaction at Each Prompt (RQ1)}
We began by examining how students engaged with the LLM during the problem-solving activity at the prompt level. The 146 transcripts contained a total of 662 prompts. 

Among those, we identified 8 distinct types of engagement with LLM. Nearly every student (\(n=135\)) requested a complete solution (Theme~5) in at least one prompt. Many students (\(n=38\)) have also asked the LLM to implement a specific feature or step from the assignment instructions (Theme~6). A number of students used the LLM to improve existing code, either their own or code generated by the LLM in earlier rounds. Some of these requests involved fixing a specific error or adding a missing component (Theme~7, \(n=20\)), while others simply asked the LLM to ``fix the code'' without detailed feedback (Theme~8, \(n=27\)). 4 students requested explanations of generated code (Theme~9), and 8 asked the LLM to synthesize previous exchanges into cohesive solutions (Theme~10). Finally, some students (\(n=14\)) used the LLM for peripheral tasks unrelated to core implementation, such as adding comments or clarifying Python concepts (Theme 11). During the interaction, some of the prompts did not initiate new requests but answered or confirmed what the LLM generated in the previous round of conversation, such as ``yes''. We categorize such prompts as ``Response to AI'' (Theme~12, \(n=24\)).

When engaging with LLM, students often provided context and instructions. We identified four types of approaches. The most common approach was to upload or copy–paste instructor-provided instructions, including the assignment description (Theme~1, \(n=92\)) and skeleton code (Theme~2, \(n=123\)). Others selectively copied relevant excerpts for a particular feature or step (Theme~3, \(n=45\)), or paraphrased the instructions in their own words (Theme~4, \(n=14\)). 

During their interaction with LLM, students usually use a mix of ways of engagement and approaches to provide contexts. The last column of table~\ref{tab:prompt-codebook} summarizes the number of prompts observed in each theme.

% \subsubsection{Codebook Introduction}
% \carol{To systematically analyze how students engaged with the LLM during problem solving, we constructed a prompt-level codebook with 12 themes (see Table~\ref{tab:prompt-codebook}). A separate code similarity codebook is introduced later in RQ2. The prompt-level themes represent distinct functional roles of prompts and were organized into three broad categories to mirror the natural flow of interaction. \textit{Context provision} (Themes 1–4) captures how students set up the problem, often by supplying instructions or template code. \textit{Task engagement} (Themes 5–11) reflects the primary forms of interaction, such as requesting complete solutions, asking for targeted features, or seeking clarification. Finally, \textit{conversational uptake} (Theme 12) marks occasions when students directly responded to the LLM’s suggestions, signaling a more dialogic exchange.} 

\begin{table}{!htb}
\centering
\begin{tabular}{@{} p{0.01\textwidth} p{0.21\textwidth} p{0.58\textwidth} p{0.10\textwidth} @{}} 
\toprule
\textbf{ID} & \textbf{Theme Name} & \textbf{Description} & \textbf{\# Prompt} \\
\midrule
\multicolumn{4}{l}{\textbf{Providing Contexts}} \\ \cmidrule(r){1-4}
1 & Assignment Instructions & Prompt uploaded/pasted the entire instruction from the Word Doc & 93 (14.0\%) \\
2 & Code Skeleton & Prompt uploaded/pasted the template code from the \texttt{.py} file & 126 (19.0\%) \\
3 & Selective Instructions & Prompt includes a screenshot or pasted text of step-wise instructions, feature descriptions, or output examples & 60 (9.1\%) \\
4 & Self-Explanation & Prompt includes the student’s own description of the entire assignment, step-wise instructions, feature descriptions, or output examples & 25 (3.8\% \\
\midrule
\multicolumn{4}{l}{\textbf{Engaging with LLM}} \\ \cmidrule(r){1-4}
5 & Complete code solution & Prompt explicitly or implicitly requests the entire code solution & 187 (28.2\%) \\
6 & Step/feature code & Prompt explicitly or implicitly requests a certain step or feature & 57 (8.6\%) \\
7 & Error/missing pieces & Prompt requests code fix or modification by describing an error or pointing out missing pieces & 27 (4.1\%) \\
8 & General fix/modification & Prompt requests code modification without specific requirements & 34 (5.2\%) \\
9 & Explanation & Prompt requests AI to explain the generated code & 4 (.6\%) \\
10 & Synthesis & Prompt requests AI to synthesize outcomes from earlier conversations & 8 (1.2\%) \\
11 & Other Guidance & Prompt does not ask for an answer or code implementation & 15 (2.3\%) \\
% \midrule
% \multicolumn{4}{l}{\textbf{Conversational Uptake}} \\ \cmidrule(r){1-4}
12 & Response to AI & Prompt answers a question from AI or accepts an AI-suggested action & 26 (3.9\%) \\
\bottomrule
\end{tabular}
\caption{Prompt-level interaction themes and definitions, with observed number of prompts and percentages of each theme across all coded student prompts (\(N = 662\)).}
\label{tab:prompt-codebook}
\end{table}
\begin{figure}{!htb}
    \centering
    \includegraphics[width=0.8\linewidth]{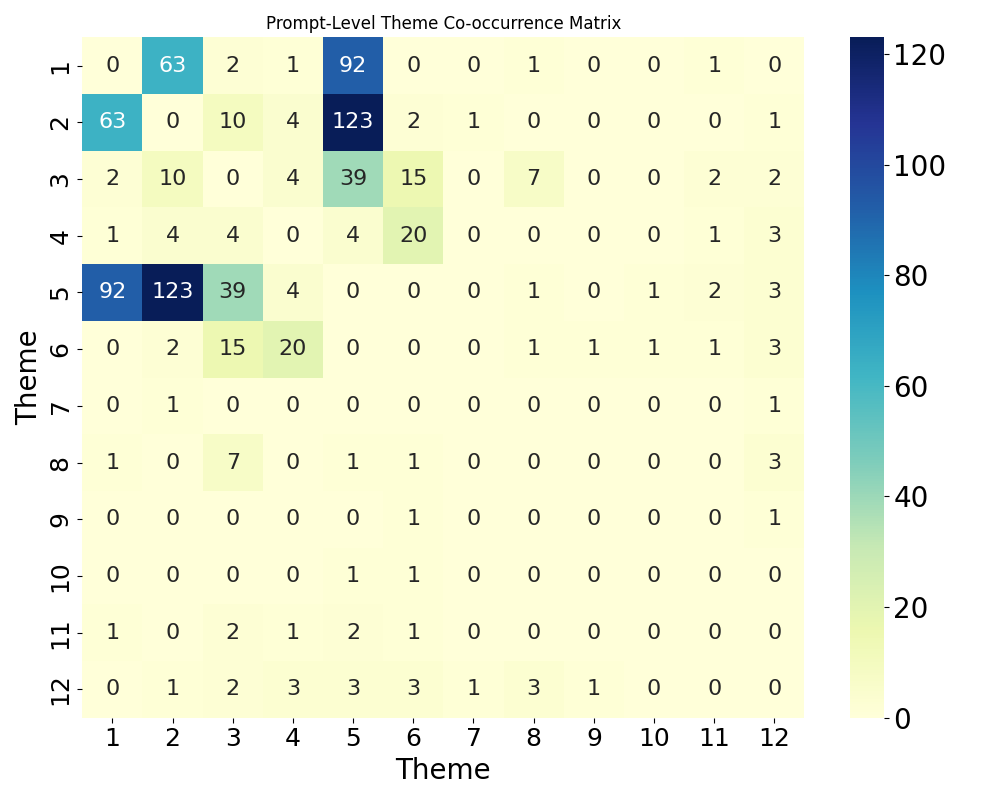}
    \caption{Prompt-level co-occurrence matrix of prompt themes. Darker cells indicate stronger co-occurrence frequencies.}
    \label{fig:Prompt-cooccurrence}
\end{figure}

We further analyzed the \textbf{co-occurrence of themes at the prompt level} to better understand how students interacted with LLM at each prompt (Figure~\ref{fig:Prompt-cooccurrence}). We found that Themes~1 and 2 (assignment instructions and skeleton code) frequently co-occurred with Theme~5 (requests for complete solutions). Theme~3 (selective instructions) often appeared alongside both Theme~5 and Theme~6 (request for stepwise or feature-wise implementations). Theme~4 (student-paraphrased instructions), although less common, was most often paired with Theme~6 when students requested partial, stepwise solutions. In other words, when asking for code solutions, whether complete or partial, most students uploaded the instructor-provided materials, while a small percentage chose to describe the tasks in their own words.

When students used the LLM to improve code, they typically provided error details or identified missing pieces directly (Theme~7), without relying on the instructor-provided materials (Themes~1-3). By contrast, when requesting a generic fix (Theme~8), they often supplied a relevant excerpt of the instructions (Theme~3), likely assuming the LLM could infer the necessary connections.  

% \carol{Round-level co-occurrence analysis further highlights these trends (Figure~\ref{fig:round-cooccurrence}). The strongest overlaps were between Theme 2 (\textit{Template code}) and Theme 5 (\textit{Complete code solution}) with 124 co-occurrences, and between Theme 1 (\textit{Full instructions}) and Theme 5 with 93 co-occurrences. These combinations suggest that students often bundled detailed context, such as assignment instructions or template code, directly with a request for a complete solution, effectively positioning the LLM as a turnkey code provider. Smaller but meaningful co-occurrences included Theme 3 (\textit{Partial instructions}) with Theme 5 (39 co-occurrences) and with Theme 6 (15 co-occurrences). By contrast, repair and reflective themes (7-12) rarely co-occurred with other themes, indicating that these behaviors typically unfolded across turns rather than within a single prompt.}

Overall, the distribution and co-occurrence analyses point to a long-tail pattern in which a small number of comprehensive solution requests dominated most student–LLM interactions, whereas repair and reflection appeared only occasionally. %Moreover, the co-occurrence patterns highlight how students frequently coupled context provision with solution requests in the same message, reinforcing prior findings that LLMs were often approached as code generators rather than as tutors or collaborative partners. 
These prompt-level findings demonstrate which prompt types were most prevalent, but they do not capture how behaviors shifted over time. To address this, we constructed interaction trajectories by concatenating the themes on prompt-level engagement (Themes~5-12) in the order of conversation rounds for each student.

\subsection{Student-AI Interaction Trajectories Throughout (Timed) Problem Solving Process (RQ1)}
Each transcript records student-LLM interaction within the 50-minute problem-solving process during the lecture session. 
Drawing on the initial engagement theme in Round~1, the overall interaction outcome (similarity between the students' submitted Python programs and the LLM's responses), and the ways subsequent prompts repeated or alternated across themes, we identified eight distinct interaction trajectories across the 146 submissions (Table~\ref{tab:trajectories}).

\begin{table}{!htb}
\centering
\begin{tabular}{@{} p{0.02\textwidth} p{0.18\textwidth} p{0.65\textwidth} p{0.09\textwidth} @{}} 
\toprule
\textbf{ID} & \textbf{Trajectory} & \textbf{Description} & \textbf{Count} \\
\midrule
% \multicolumn{4}{l}{\textbf{Hybrid Trajectories (Section~4.3)}} \\ \cmidrule(r){1-4}
1 & Simple Delegation & Students leveraged the LLM to generate a complete solution in a single step, aiming for efficiency and full-task resolution. & 52 (35.6\%) \\
2 & Worked-Example Adaptation & Students obtained AI-generated code once, then adapted or debugged it to produce a noticeably modified submission. & 7 (4.8\%) \\
3 & Minimal Use & Students interacted briefly with the LLM but ultimately relied on their own coding, producing final work with little overlap with AI output. & 10 (6.8\%) \\
% \midrule
% \multicolumn{4}{l}{\textbf{Multi-round Prompt-based Trajectories (Section~4.1.4)}} \\ \cmidrule(r){1-4}
4 & Persistent Delegation & Students engaged the LLM across multiple rounds to address the entire problem, re-asking or rephrasing when uncertain about how to begin or proceed. & 37 (25.0\%) \\
5 & Stepwise Exploration & Students decomposed the task into smaller features and were prompted sequentially for each component. & 5 (3.4\%) \\
6 & Backwards Scaffolding & Students alternated between requesting code and debugging it, creating iterative loops of generation and repair. & 35 (23.6\%) \\
7 & Debugging Collaboration & Students engaged the LLM primarily as a repair partner, opening with or focusing exclusively on debugging prompts. & 3 (2.0\%) \\
8 & Cycle of Dependency & Students cycled between generation and repair but became locked in repeated fix loops without stabilizing a final solution. & 8 (5.4\%) \\
\bottomrule
\end{tabular}
\caption{Interaction trajectory descriptions, and observed distributions.}
\label{tab:trajectories}
\end{table}

% Some interacted briefly, obtaining a solution in one step, while others engaged in extended sessions that mixed comprehensive requests with cycles of repair or incremental refinement. These patterns also point to differences in how much cognitive effort students invested: some offloaded most of the work to the LLM, while others devoted more energy to adapting, repairing, or extending its outputs. Taken together, prompt types and trajectories provide insight into the strategies students adopted and the roles they assigned to the AI during the assignment. In the subsections that follow, we introduce the codebooks that guided our analysis, present the distribution and temporal progression of prompt types, and then describe the major trajectories that emerged. Analyses of code similarity are presented in RQ2, as here we focus specifically on prompting behaviors and their evolution.
% \subsubsection{Temporal Evolution of Prompts}
% \carol{Although all 12 coded themes describe aspects of student–LLM interactions, our analysis focuses on Themes 5–12 because they show how students actively engaged with the system. Themes 1–4 provided task context, such as pasting instructions or code templates, but they did not reveal students’ orientations toward the LLM. Themes 5–12, in contrast, capture how students sought complete solutions, attempted to fix problems, or pursued clarification.}
Among the 146 valid submissions, 59 (40.4\%) students only had one round of conversation with LLM, i.e., one student prompt, one LLM response. Of these, the majority of students (\(n=55 \)) requested a complete solution to the entire assignment, while very few students (\(n=4\)) prompted other themes [themes 8 (\(n=2\)) and 11 (\(n=2\))]. When comparing the LLM responses with the students' final submissions, we observed that 53 students directly submitted LLM-generated code or pasted step by step. Only one student made modifications based on LLM-generated code. Six students submitted code that was completely different from the LLM responses. Table~\ref{tab:rounds-distribution} summarizes the distribution of conversation rounds.

Overall, we synthesized eight interaction trajectories across the valid submissions.

\paragraph{Trj 1. Simple Delegation} 
There are 52 students (35.6\%) who used LLM to generate a complete solution to the assignment with one prompt and submitted the LLM-generated code. These students delegated the entire problem-solving process to the LLM and did not modify LLM-generated solutions. There is no evidence to show that the students reviewed and understood the LLM-generated code. %They decided that they fully agree with the solution. 

\paragraph{Trj 2. Worked-Example Adaptation}
In contrast, seven students who also used the LLM to produce a complete solution with one prompt made substantive modifications before submitting their work. These changes went beyond cosmetic edits or simple refactoring and involved structural adjustments such as modifying loops or conditional statements. This suggests that these students examined and tested the LLM-generated code in light of the assignment requirements and applied what they learned to meaningfully improve the solution.

\paragraph{Trj 3. Minimal Use}
Notably, 10 students submitted code that was entirely different from the AI-generated solutions. One possibility is that their interaction with the LLM was primarily driven by the course activity requirements, and that they intentionally chose not to rely on the LLM when solving the assignment. Alternatively, these students may have considered AI-generated solutions as ineffective and chose to construct their own solutions instead of revising what AI suggested.

\paragraph{Trj 4. Persistent Delegation}
Thirty seven students repeatedly requested the LLM to produce complete solutions to the assignment (Theme~5) across the entire interaction, with prompts of different rounds varying only in the type of context or instructions they provided (Themes~1–4). At times, they requested modifications without offering specific feedback on what needed improvement (Theme~8), and/or requested synthesis of AI-generated code and requested modifications (Theme~10). Throughout the interaction, the students primarily changed the way of giving instructions for LLM to solve the entire assignment on their behalf. %Therefore, we classify such interaction trajectories as persistent delegation.

\begin{figure}[!htb]
    \centering
    \includegraphics[width=\linewidth]{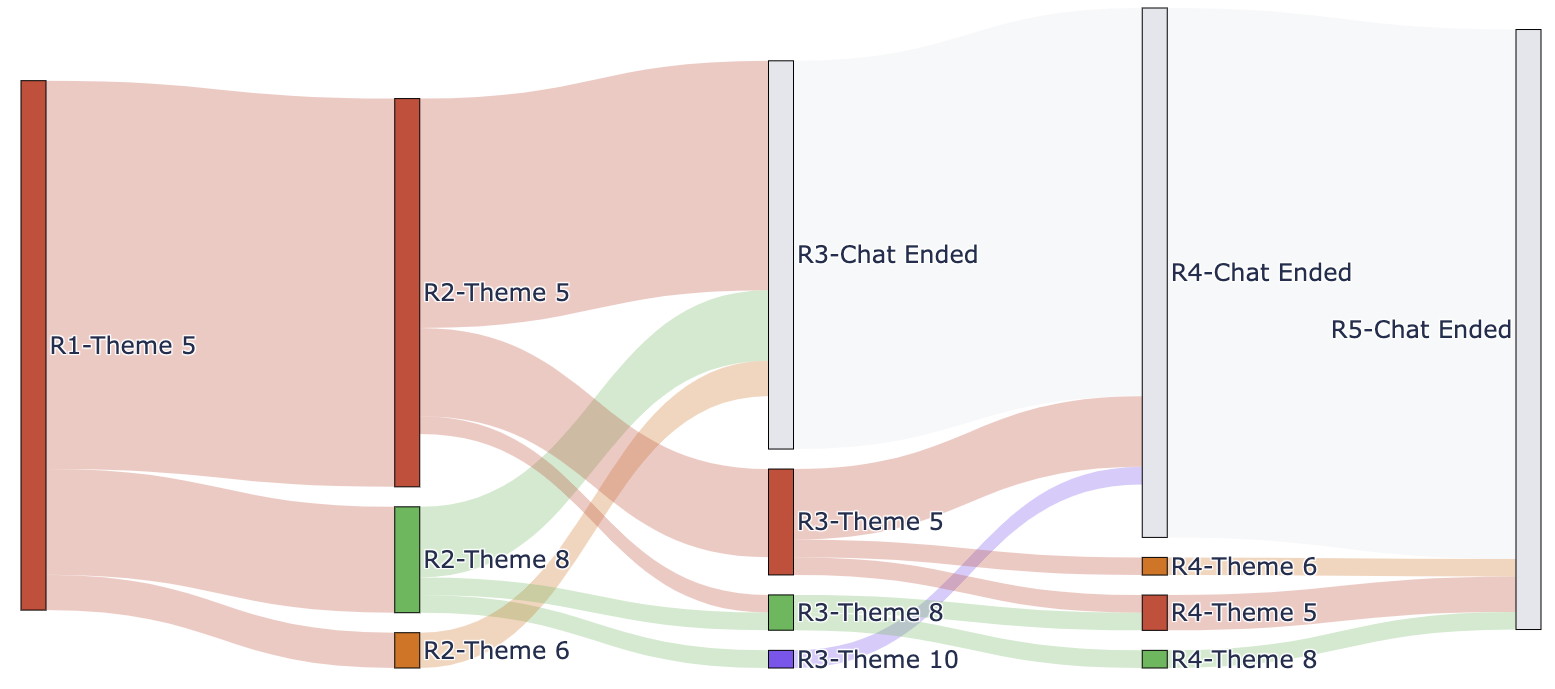}
    \caption{Sankey diagram of student–LLM conversation flows that follows the \textbf{Persistent Delegation} trajectory. It portrays that students repeatedly requested complete solutions (Theme~5) across multiple rounds, occasionally supplementing with generic fix requests (Theme~8) or stepwise prompts (Theme~6), but without substantially changing their strategy. Most conversations ended after a few rounds, with students continuing to rely on the LLM for complete code rather than engaging in iterative refinement.}
    \label{fig:persistent-delegation}
\end{figure}

\paragraph{Trj 5.Stepwise Exploration}
This trajectory reflects students’ use of the LLM to support problem-solving step by step, following the features and tasks defined in the assignment instructions. Students typically began by asking the LLM to implement a single step, then progressively requested additional steps in sequence (Theme~6). In some cases, the interaction concluded with a request for a generic code improvement (Theme~8). Overall, students leveraged the instructor-provided scaffolding to structure the LLM’s contributions, while still evaluating the implementation of each step themselves. %We therefore classify this interaction pattern as stepwise exploration.

\begin{figure}[!htb]
    \centering
    \includegraphics[width=\linewidth]{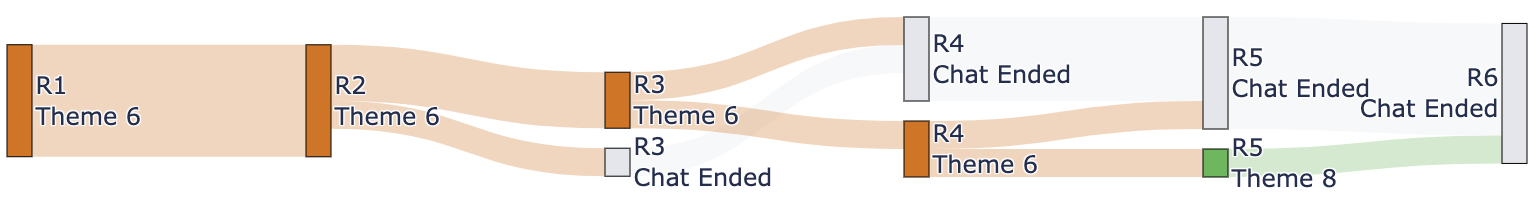}
    \caption{Sankey diagram of student–LLM conversation flows that follows the \textbf{Stepwise Exploration} trajectory. It portrays that students began by requesting stepwise or feature-specific code implementations (Theme~6) and continued to prompt the LLM for additional components across subsequent rounds. Most conversations ended after several iterations, while a few transitioned into generic fix requests (Theme~8) before concluding.}
    \label{fig:stepwise-exploration}
\end{figure}

\paragraph{Trj 6. Backwards Scaffolding}
Rather than repeating single requests, many students began by asking the LLM for a complete solution (Theme~5) and then iteratively revised that solution through subsequent prompts. These revisions included providing feedback on errors or missing pieces (Theme~7), guiding the LLM to regenerate components for specific steps or features (Theme~6), occasionally making more general requests for code fixes (Theme~8), and synthesizing the outcome into a cohesive code file (Theme~10). This iterative cycle of generation and repair illustrates how students initially delegated problem-solving to the LLM but later collaboratively scaffolded the LLM to refine the solution. %We refer to this type of interaction trajectory as backwards scaffolding.

\begin{figure}[!htb]
    \centering
    \includegraphics[width=\linewidth]{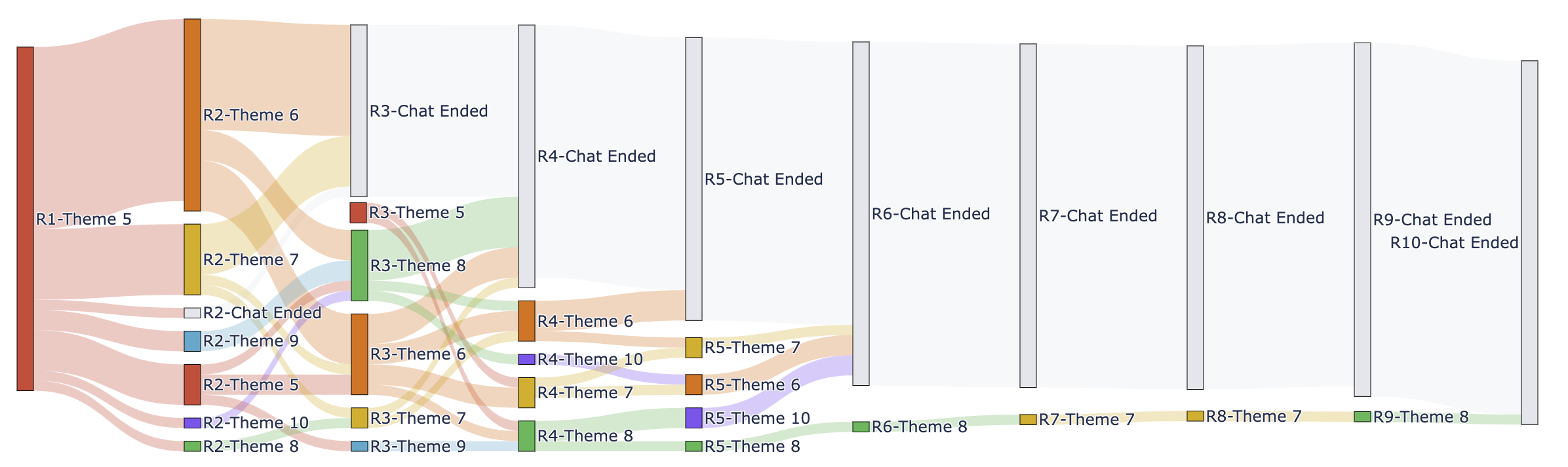}
    \caption{Sankey diagram of student–LLM conversation flows that follows the \textbf{Backwards Scaffolding} trajectory. It portrays that students began by requesting complete solutions (Theme~5) in the early rounds, then transitioned into stepwise feature requests (Theme~6), debugging prompts (Theme~7), generic fixes (Theme~8), or synthesis (Theme~10).}
    \label{fig:backwards-scaffolding}
\end{figure}

\paragraph{Trj 7. Debugging Collaboration} 
Three students uploaded their own code and asked the LLM to help with debugging specific errors (Theme~7) or improve the student solution based on the assignment instruction (Theme~5 and 8). 

\paragraph{Trj 8. Cycle of Dependency}
Some students combined elements of Backwards Scaffolding and Debugging Collaboration in their interactions with the LLM, alternating between requests for complete or partial solutions (Themes~5 and~6) and debugging or improvement prompts (Themes~7 and~8). In a few cases, students also synthesized outcomes from earlier conversations before continuing to refine the solution with the LLM. These interactions typically oscillated between generating new code and fixing the resulting output from one prompt to the next. %We therefore classify such trajectories as a cycle of dependency. 

\begin{figure}[!htb]
    \centering
    \includegraphics[width=\linewidth]{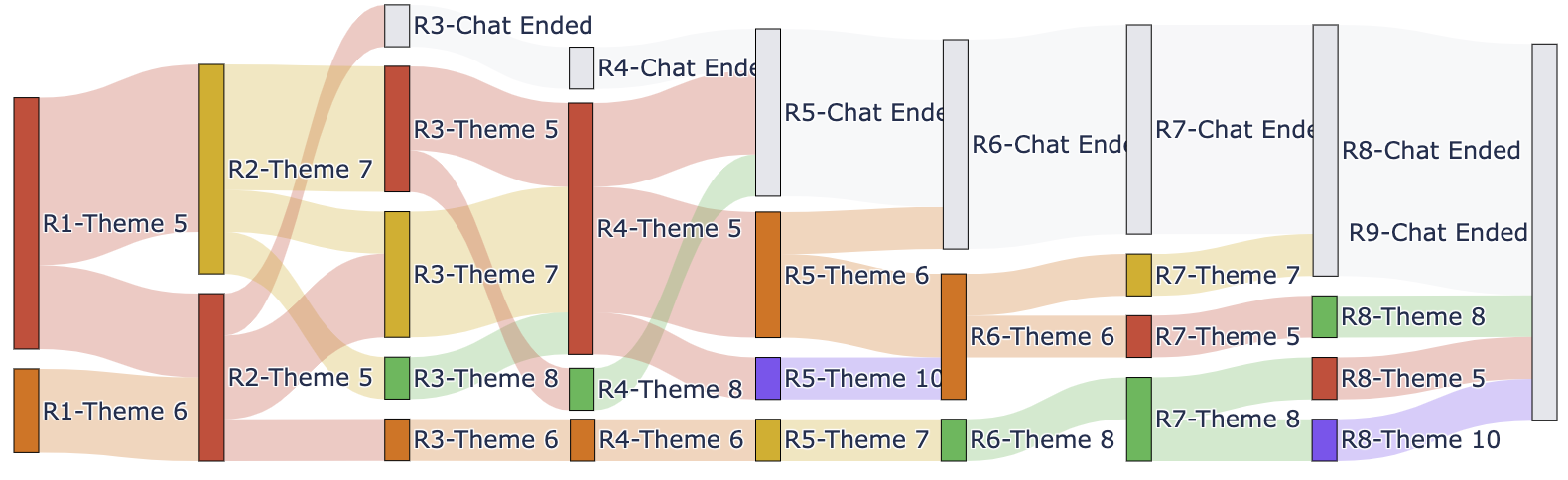}
    \caption{Sankey diagram of student–LLM conversation flows that follows the \textbf{Cycle of Dependency} trajectory. Students began by requesting either complete solutions (Theme~5) or stepwise implementations (Theme~6) in the first one or two rounds, then shifted into repeated debugging interactions (Themes~7–8). Rather than stabilizing on a working solution, these conversations alternate between requesting new solutions and further fixes. }
    \label{fig:backwards-scaffolding}
\end{figure}
% \carol{Five main multi-round trajectories emerged (Table~\ref{tab:trajectories}, top half). \textit{Comprehensive Problem Navigation} ($n = 37$, 25.0\%) described students who returned to complete solutions in multiple rounds, often using them as anchor points when navigating the assignment as a whole. This approach reflected a reliance on full examples to maintain orientation and ensure coverage of the task. \textit{Stepwise Exploration} ($n = 5$, 3.4\%) was less common and reflected an attempt to break the assignment into features, prompting for one component at a time. \textit{Backwards Scaffolding} ($n = 35$, 23.6\%) described a back-and-forth pattern in which students alternated between generating new code and asking the LLM to correct problems in earlier outputs. \textit{Debugging Collaboration} ($n = 3$, 2.0\%) was defined by students who treated the LLM mainly as a repair partner, opening with or focusing on debugging requests. \textit{Cycle of Dependency} ($n = 8$, 5.4\%) captured situations where students fell into repeated fix loops and continuously prompted the AI for corrections without arriving at a working solution. Alongside these five categories, a few special cases were observed. For example, one student consistently asked the LLM to explain code rather than produce it, which stood out as an outlier trajectory. The hybrid trajectories combine prompting behaviors with the extent of code reuse and are discussed in Section~4.3.}

\subsection{Code Similarity and Convergence (RQ2)}
We analyzed similarity in students' Python code through two complementary lenses. First, we qualitatively labeled the degree of overlap between the student's submitted Python file and the code generated during their interaction with LLMs. This includes comparing all the code snippets generated in the chat history with the final implementation in the Python file. %This coding captures surface-level and structural resemblances that reflect how closely students followed LLM-generated outputs. 
Second, we quantitatively clustered all Python submissions based on rubric-based, structural, and semantic features. %These analyses allowed us to test whether students gravitated toward a small number of solution types or distributed more evenly across diverse styles.
\begin{table}[!htb]
\centering
\begin{tabular}{@{} p{0.01\textwidth} p{0.2\textwidth} p{0.52\textwidth} p{0.12\textwidth} @{}} 
\toprule
\textbf{ID} & \textbf{Theme Name} & \textbf{Description} & \textbf{\# Students} \\
\midrule
1 & Direct Copy & Verbatim reuse of AI-generated code in the final \texttt{.py} submission. & 119 (80.4\%) \\
2 & Minor Modification & Small, surface-level edits to AI code (e.g., variable renaming, formatting, comment changes). & 6 (4.0\%) \\
3 & Iterative Incorporation & AI code reused with noticeable debugging, integration, or targeted fixes across iterations. & 5 (3.4\%) \\
4 & Substantial Modification & AI code adapted heavily with added or restructured logic; clear author changes beyond surface edits. & 6 (4.0\%) \\
5 & Independent Code & Final submission diverges significantly from AI outputs; minimal or no reuse evident. & 10 (6.8\%) \\
\bottomrule
\end{tabular}
\caption{Code similarity codebook with observed frequencies and percentages of each theme across all final student submissions.}
\label{tab:code-similarity}
\end{table}
\subsubsection{Most Students Included AI-Generated Code in their Assignment Submission}
Our qualitative analysis identified five levels of similarity between the AI-generated code and the students' assignment submission, ranging from exactly the same (direct copy) to completely different (independent code), see Table~\ref{tab:code-similarity}. %Each submission was manually labeled according to one of these five levels.
%To examine how AI outputs influenced students’ final submissions, we applied the code similarity codebook to all 148 student programs. Each submission was labeled into one of five categories representing the degree of overlap with LLM-generated code (Table~\ref{tab:code-similarity}). 
The results show a strong tendency toward direct reuse: nearly three-quarters of students (84.4\%, \(n=125\)) submitted code that was a near-verbatim copy of the LLM’s output, with at most cosmetic edits or minor refactoring. Five students incorporated LLM-generated code with limited but meaningful modifications, such as addressing minor bugs. 6 students made more substantial alterations, including restructuring program logic or adding original components. 10 students (6.8\%) produced independent code that showed little to no overlap with the AI outputs. %An additional 1.4\% ($n = 2$) were coded as \textit{Other}, representing outlier cases that did not fit cleanly into the five-level spectrum (e.g., corrupted files or incomplete submissions).

\subsubsection{Students' Submitted Code Mostly Converge into Adjacent Clusters}
% To assess convergence and divergence across students, we applied k-means clustering to the 146 Python submissions using three types of feature sets: traditional structural features, and semantic embeddings from CodeBERT~\cite{feng2020codebert}\footnote{Embeddings are vector representations of code that capture semantic similarity. They allow machine learning models to compare and cluster submissions not only
% by syntax but also by meaning. }. We also examined combined feature sets to evaluate whether structural and semantic information together produced more meaningful groupings. %Cluster quality was evaluated using the Silhouette score~\cite{}, and elbow plots guided the selection of the optimal number of clusters.

The rubric-based clustering produced four clusters with a moderate Silhouette score~\cite{rousseeuw1987silhouettes} (0.38)\footnote{Silhouette values are typically interpreted as follows: \((s < 0.25)\): the clustering structure is very weak or artificial; \((0.25 \leq s < 0.50)\): clustering suggests potentially meaningful structure but with substantial overlap; \((0.50 \leq s < 0.70)\): clustering indicates reasonable structure; \((s \geq 0.70)\): clustering shows strong and well-separated structure~\cite{rousseeuw1987silhouettes}.}, indicating potentially meaningful but low separation. %In this sense, our observed silhouette scores fall within a range where clusters can still be interpreted as meaningful, even if separation is limited. Most students fell into a large middle-performing cluster (58.9\%), with a smaller but substantial group showing higher performance (36.4\%), and two very small clusters containing students whose code had runtime errors (4.6\% combined). Although the Silhouette values indicate only modest cohesion, these clusters remain educationally meaningful because they directly reflect observable differences in code correctness and performance.
Clustering on traditional structural features produced three clusters with a higher Silhouette score (0.46), though this result was likely influenced by redundancy among correlated features (e.g., line counts and character counts). After filtering to reduce multicollinearity, the optimal solution expanded to six clusters, but with a much lower Silhouette score (0.22), suggesting weaker separation.
Semantic clustering with CodeBERT embeddings produced four clusters, with two large groups ($\sim$45\% each) and two very small groups ($\sim$5\% each). Retaining comments slightly improved cohesion (Silhouette=0.183 vs.\ 0.152 without comments), though both scores were low. When combined with traditional features, clustering again produced four groups, but with similarly weak Silhouette values (0.142--0.179).

The generally low Silhouette scores across feature sets are not unexpected, as all students were provided with the same code skeleton that constrained variation in program structure and logic. Even so, a consistent pattern emerged across the three different methods: one large cluster representing the majority of students who converged on similar implementations, accompanied by smaller clusters that reflected higher-performing, divergent, or error-prone solutions. These findings are aligned with the qualitative analysis results (Section~\ref{tab:code-similarity}), where most students heavily reused LLM-generated code with only a few making meaningful modifications or creating independent solutions. %Together, both analyses suggest that while the template and assignment design promoted convergence, meaningful variation persisted at the margins.

% \subsubsection{Synthesis and Implications}
% Both the code similarity and clustering analyses point to strong convergence in student work. Most submissions closely mirrored LLM outputs and clustered into a dominant group of similar implementations, while only a minority showed substantial modification or independent solutions. The presence of smaller clusters and outlier cases highlights that some students invested more cognitive effort than others, creating variation at the margins. These outcome patterns frame RQ3, where we examine how interaction trajectories explain differences in code reliance and performance.

\subsection{Assignment and Course Performance of Different Interaction Orientations (RQ3)}

To examine how student-LLM interaction styles may relate to both assignment-level and overall course performance, we compared rubric-based assignment scores and final letter grades across the eight trajectories. Assignment scores were evaluated by the course teaching assistant using a rubric (Table~\ref{tab:code-rubric}) that reflects both correctness and adherence to assignment instructions. Total points evaluate to 130. Final letter grades were determined by the course instructor based on all graded components, following the predefined grading formula and scheme.

\begin{figure}
    \centering
    \includegraphics[width=\linewidth]{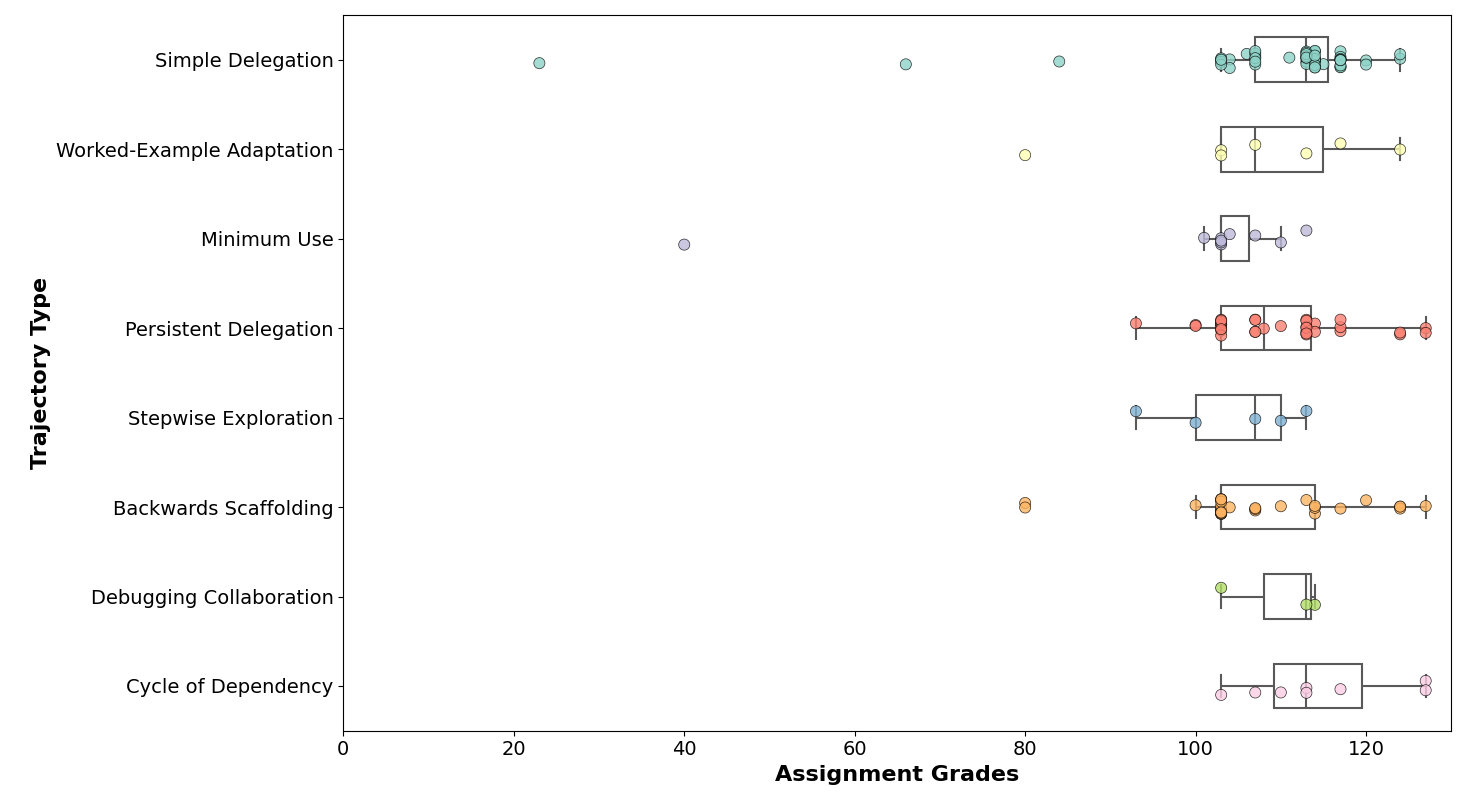}
    \caption{Assignment grades across different LLM interaction trajectories. }
    \label{fig:assignment-grade}
\end{figure}

The overall assignment performance ranged from 23 to 127 points (\(M=108.29\), \(Mdn = 110\), \(SD = 12.79\)). To examine whether assignment performance differed across interaction trajectories, we conducted a Kruskal–Wallis test~\cite{kruskal1952use} on assignment grades grouped by trajectory type. Results revealed a significant overall difference among the 12 trajectory categories (\(H(11)=19.83\), \(p=0.01\)).
Post-hoc pairwise comparisons using Mann–Whitney U tests~\cite{mann1947test} with Holm correction~\cite{holm1979simple} revealed that students in the \textit{Minimal Use} trajectory scored significantly lower than those in the \textit{Simple Delegation} trajectory (p < .05). Specifically, the median grade for Minimum Use students was 103 (\(M=97.4\)), compared to a median of 113 (\(M= 109.0\)) for Simple Delegation students. No other pairwise differences were statistically significant after correction.
This contrast between Minimum Use and Simple Delegation is particularly noteworthy because the two represent opposite interaction styles. Students who did not use the LLM scored significantly lower than those who fully outsourced their work to it. However, overall course grades did not differ significantly between these two groups. Among students in the Simple Delegation trajectory (\(n=50\)), 29 received an A- to A+ grade, 20 received a B- to B+, with 1 receiving a D. In contrast, students in the Minimum Use trajectory (\(n=9\)) 6 received A- to A+, and the other 3 received B+. 

To assess whether overall course performance differed across trajectories, we treated students' final letter grades as ordinal data (A+ > A > A- > … > F) and conducted a Kruskal–Wallis test~\cite{kruskal1952use}. Median grades clustered around A– to A for Stepwise Exploration, Persistent Delegation, Backwards Scaffolding, and Simple Delegation, while Cycle of Dependency tended to be slightly lower (median B+). The Kruskal–Wallis test revealed no significant differences among trajectories, \(H(5)=4.59\), \(p=.47\). Thus, overall grade performance did not significantly vary by trajectory types.

% \subsubsection{Synthesis and Implications}
% The trajectory analysis shows that interaction styles strongly influenced both code reuse and assignment outcomes. End-to-End Problem Solving and Comprehensive Problem Navigation were the most frequent strategies, and both often produced stable submissions within the 50-minute limit. Cyclical approaches, such as the Cycle of Dependency, also yielded strong results, most likely because repeated requests and fixes created additional chances to correct errors. In contrast, the Minimum Use group underperformed relative to most others. Their independence reduced access to LLM support, which left many unable to meet rubric expectations. Adaptation-based paths produced mixed results. For some students, modification of the LLM code improved quality, while for others, the extra effort consumed valuable time without adding correctness.

% These patterns suggest that strategies that minimized cognitive burden, either by reusing solutions directly or by cycling through fixes, were well-suited to the time-limited setting. Independent paths demanded more sustained effort and carried higher risks, resulting in uneven outcomes. This tension between efficiency and effort highlights how students made deliberate trade-offs in allocating cognitive resources. The following Discussion section returns to this theme of cognitive load and explores its implications for designing educational activities with LLMs.
\section{Discussion}
\subsection{Interaction Trajectories as Cognitive Offloading Strategies}
Our analysis shows that student–LLM interaction trajectories can be understood as distinct strategies for managing cognitive load under time pressure. Simple Delegation minimized both \textit{intrinsic} and \textit{extraneous} load by outsourcing the entire task to the LLM, though often at the cost of shallow engagement. Worked-Example Adaptation, by contrast, illustrates how learners offloaded low-level construction while retaining \textit{germane load} for evaluating and refining the example, consistent with cognitive load theory’s emphasis on worked examples~\cite{van2010cognitive}. Backwards Scaffolding and Debugging Collaboration further highlight how students distributed effort across human and AI: the LLM absorbed the cost of generating and repairing code, while students focused on directing and testing corrections. These strategies lowered the barrier to engaging in productive failure by making trial-and-error cycles less costly.

From this perspective, the LLM acted as a flexible cognitive partner—sometimes as a solution provider, other times as a debugging collaborator. Which role it played depended less on the system itself than on students’ orientations and choices within a constrained, time-limited task. Even in trajectories of persistent delegation, students were not entirely passive: they reformulated instructions, experimented with different ways of supplying context, and likely tested outputs against requirements. This indicates that offloading to the LLM shifts rather than eliminates effort, redistributing it from code construction to interaction management and evaluation, moving up in the bloom's pyramid~\cite{bloom1964taxonomy}.

These findings suggest important implications for instructional design. If left unstructured, many students will default to efficiency-seeking trajectories that maximize output but minimize reflection. Educators can counterbalance this tendency by designing tasks that explicitly reward evaluation and adaptation of AI outputs, for example by requiring reflection on modifications, structuring assignments into phases that separate independent attempts from LLM use, or assessing not only correctness but also evidence of testing and revision. Such scaffolds can guide students toward trajectories where the LLM supports—not supplants—learning.

\subsection{Convergence and Homogeneity in Student Code}
A common concern in LLM-supported programming is whether students’ submissions converge toward homogeneous solutions~\cite{amoozadeh2024student, budhiraja2024its, jost2024impact}. %Heavy reliance on generated code could narrow the diversity of approaches, while varied prompting and integration strategies might preserve differences in structure and style.
Our analyses revealed strong convergence. Over 80\% of students directly reused AI-generated code with little or no modification, and clustering across rubric-based, structural, and semantic features consistently yielded a small number of clusters. Silhouette scores were uniformly low, reflecting limited separation and indicating that students’ final implementations were highly similar.

This homogeneity, however, is partly by design. The assignment provided a code skeleton and explicit requirements, which constrained the solution space. LLM reuse reinforced shared patterns. In this sense, convergence is not necessarily problematic: for students still acquiring core programming skills, arriving at functionally correct and structurally consistent solutions can reinforce syntax, control flow, and confidence. Convergence may even signal that the LLM effectively helped most students reach a baseline solution.

At the same time, variation persisted at the margins. A small subset of students substantially adapted or extended LLM outputs, while others diverged completely, producing independent solutions or struggling to stabilize a working program. These cases highlight that while convergence streamlined progress, opportunities for deeper engagement and creativity still existed.

Overall, convergence in this context reflects how LLMs, combined with structured assignments, can efficiently scaffold foundational learning. The pedagogical challenge is to decide when uniformity is desirable for reinforcing basics and when assignments should be designed to encourage diversity of approaches, adaptation of examples, or exploration beyond the template.

\subsection{Assessment and Equity Considerations}
Our findings raise important questions about how to fairly assess student work in the presence of LLM support. Trajectories such as Minimal Use and Simple Delegation produced significantly different assignment outcomes, yet overall course performance did not differ between these groups. This suggests that performance in a single timed activity is not predictive of longer-term learning or achievement. %Over-reliance on assignment-level outcomes could therefore disadvantage students who, by choice or by skill, engage with LLMs differently.

Prior research has highlighted disparities in prompt-crafting ability, and our observations extend this concern~\cite{kumar2024guiding, nguyen2024how, kazemitabaar2024how}. By Spring 2025, most students behaved as though they already knew what to expect from an LLM, treating it as a familiar partner rather than an unfamiliar or rule-based system. Only one or two students attempted to interact as if the LLM operated deterministically (e.g., asking whether the LLM could execute functions). More commonly, students uploaded the full assignment instructions with minimal elaboration, relying on the assumption that the LLM would ``understand'' the context. This efficiency-oriented strategy reflects an implicit mental model shaped by prior experience. While effective for producing runnable code, it bypasses opportunities for students to articulate requirements in their own words—a skill central to problem-solving competence.

These patterns point to an emerging baseline: students increasingly bring prior expectations, shortcuts, and strategies of using LLMs into the classroom. The challenge for educators is less about teaching how to use an LLM and more about addressing how prior habits bias integration into coursework. Designing interventions that require paraphrased instructions, reflective explanations, or explicit justification of modifications can surface and expand prompting practices, turning efficiency-driven strategies into deeper learning opportunities. Without such scaffolding, AI integration risks amplifying inequities by rewarding those already adept at leveraging the tool. %Institutions will therefore need to rethink evaluation paradigms so that assessments capture not only correctness, but also process, reasoning, and growth.

\subsection{Design Implications: Learning as Prototyping}
Taken together, these findings suggest that students increasingly engage with programming assignments not as linear exercises in code construction but as cycles of \textit{product prototyping} with the LLM as a collaborator.

\textbf{Scaffold prototyping practices.} For students who treat LLM outputs as draft prototypes, assignments should explicitly reward iteration, testing, and revision. Rubrics can assess not only correctness but also how students experimented with AI-generated code, documented changes, and explained trade-offs.

\textbf{Leverage worked examples as starting prototypes.} Many students used LLMs as worked-example generators. Instructors can harness this by requiring students to annotate why they modified (or retained) certain parts of the AI’s output, reframing adaptation as an intentional learning step.

\textbf{Encourage productive failure through rapid iteration.} Trajectories such as backward scaffolding and debugging collaboration show how LLMs lower the cost of productive failures by offering immediate feedback and alternative drafts. Assignments can embed cycles of failure and recovery, positioning mistakes as integral to the prototyping process.

\textbf{Surface prompting as part of the prototype.} Many students defaulted to uploading instructions verbatim, reflecting a mental model of the LLM as an ``all-knowing interpreter.'' Designing activities that require paraphrased instructions, structured decomposition, or justification of prompts reframes prompting itself as a critical layer of the prototyping cycle.

% \textbf{Balance convergence and creativity.} Homogeneity may be pedagogically useful in early assignments, ensuring that students converge on working prototypes and gain confidence with syntax and structure. Later tasks, however, should broaden the solution space, encouraging diverse approaches, creative designs, and open-ended extensions.

\textbf{Support equity in prototyping fluency.} Students bring unequal prior experiences with LLMs, shaping how fluently they prototype. Instructors can mitigate inequities by making effective practices explicit—modeling prompts, scaffolding reflection, and rewarding iteration over delegation—so that prior familiarity does not become the primary driver of success.

\subsection{Limitations and Future Directions}
First, this study examined a single assignment within a specific course context, using a provided code skeleton and tightly defined requirements. These constraints funneled students toward similar solutions and may have amplified convergence. While this design made it easier to observe trajectories, it limits generalizability to more open-ended projects where creativity plays a larger role.

Second, the activity was conducted under strict time constraints during class. Such conditions likely encouraged efficiency-seeking behaviors and shortened interaction cycles. At the same time, this mirrors a common feature of students’ authentic experience: working under deadlines and engaging in last-minute problem-solving. Studying this setting is therefore valuable for understanding how students deploy LLMs when efficiency and completion are paramount, even if it does not capture the full range of strategies visible in longer-term projects.

Third, we had originally designed a follow-up reflection assignment to capture students’ reasoning about their use of LLM in the timed problem-solving experience, but due to course scheduling constraints, it was not administered. As a result, our analysis could not directly probe students’ metacognitive processes, and their perceptions of how LLM influenced their problem-solving processes. Nevertheless, this does not diminish the value of the study. By focusing on authentic interaction traces and final code submissions, our work reveals how students use LLMs in coursework without scaffolds on when and how to use AI.

Finally, findings are tied to the capabilities of a specific LLM version used in Spring 2025. As models evolve, error profiles, responsiveness, and prompting dynamics may shift, potentially altering both the prevalence and effectiveness of the trajectories we observed. Replication with newer models, different student populations and course contexts will be necessary to assess the stability and evolution of these interaction patterns.

\subsection*{Acknowledgments of the Use of AI}
During the data analysis of this work, we piloted LLM-assisted annotation using ChatGPT to apply our prompt-level codebook to student submissions. These annotations were compared against manual coding to assess reliability and feasibility. Final coding decisions were made by human researchers. During the preparation of this manuscript, we used OpenAI’s ChatGPT (GPT-5, September 2025 release) to assist with refining writing clarity. All AI-generated text was critically reviewed, revised, and verified by the authors to ensure accuracy and appropriateness. Responsibility for the content of this article rests entirely with the authors.

\bibliographystyle{ACM-Reference-Format}
\bibliography{ref}

@article{gwet2001handbook,
  title={Handbook of inter-rater reliability},
  author={Gwet, Kilem},
  journal={Gaithersburg, MD: STATAXIS Publishing Company},
  pages={223--246},
  year={2001}
}

@article{rousseeuw1987silhouettes,
  title = {Silhouettes: A graphical aid to the interpretation and validation of cluster analysis},
  author = {Rousseeuw, Peter J},
  journal = {Journal of Computational and Applied Mathematics},
  volume = {20},
  pages = {53--65},
  year = {1987},
  publisher = {Elsevier}
}

@article{hayes2007answering,
  title = {Answering the Call for a Standard Reliability Measure for Coding Data},
  author = {Hayes, Andrew F and Krippendorff, Klaus},
  journal = {Communication Methods and Measures},
  volume = {1},
  number = {1},
  pages = {77--89},
  year = {2007},
  publisher = {Taylor & Francis},
  doi = {10.1080/19312450709336664}
}

@article{marzi2024k,
  title = {K-Alpha Calculator–Krippendorff's Alpha Calculator: A user-friendly tool for computing Krippendorff's Alpha inter-rater reliability coefficient},
  author = {Marzi, Giacomo and Balzano, Marco and Marchiori, Davide},
  journal = {MethodsX},
  volume = {12},
  pages = {102545},
  year = {2024},
  issn = {2215-0161},
  doi = {10.1016/j.mex.2023.102541}
}

@article{wongpakaran2013comparison,
  title = {A comparison of Cohen's Kappa and Gwet's AC1 when calculating inter-rater reliability coefficients: a study conducted with personality disorder samples},
  author = {Wongpakaran, Nahathai and Wongpakaran, Tinakon and Wedding, Danny and Gwet, Kilem L},
  journal = {BMC Medical Research Methodology},
  volume = {13},
  pages = {61},
  year = {2013},
  publisher = {BioMed Central},
  doi = {10.1186/1471-2288-13-61}
}

@article{ali2023assessment,
  title = {Assessment of ChatGPT-generated programming code based on exercises in an introductory programming course},
  author = {Ali, A and Wibowo, K},
  journal = {Issues in Information Systems},
  volume = {24},
  number = {2},
  pages = {203--212},
  year = {2023},
  doi = {10.48009/2_iis_2023_117}
}

@article{agbo2025computing,
  title = {Computing education using generative artificial intelligence tools: A systematic literature review},
  author = {Agbo, Friday Joseph and Olivia, Chris and Oguibe, Godsalvation and Sanusi, Ismaila Temitayo and Sani, Godwin},
  journal = {Computers and Education Open},
  volume = {9},
  pages = {100266},
  year = {2025},
  publisher = {Elsevier},
  doi = {10.1016/j.caeo.2025.100266}
}

@inproceedings{ambikairajah2024chatgpt,
  title = {ChatGPT in the Classroom: A Shift in Engineering Design Education},
  author = {Ambikairajah, E and Sirojan, T and Thiruvaran, T and Sethu, V},
  booktitle = {2024 IEEE Global Engineering Education Conference (EDUCON)},
  year = {2024},
  pages = {1--5},
  doi = {10.1109/EDUCON60312.2024.10578884},
  publisher = {IEEE}
}

@inproceedings{amoozadeh2024student,
  title = {Student-AI Interaction: A Case Study of CS1 students},
  author = {Amoozadeh, Matin and Nam, Daye and Prol, Daniel and Alfageeh, Ali and Prather, James and Hilton, Michael and Ragavan, Sruti S and Alipour, Amin},
  booktitle = {Proceedings of the 24th Koli Calling International Conference on Computing Education Research (Koli Calling '24)},
  year = {2024},
  publisher = {Association for Computing Machinery},
  pages = {1--13},
  doi = {10.1145/3699538.3699567}
}

@article{arantes2024understanding,
  title = {Understanding Intersections Between GenAI and Pre-Service Teacher Education: What Do We Need to Understand About the Changing Face of Truth in Science Education?},
  author = {Arantes, Joana},
  journal = {Journal of Science Education and Technology},
  year = {2024},
  doi = {10.1007/s10956-024-10189-7},
  publisher = {Springer}
}

@article{barambones2024chatgpt,
  title = {ChatGPT for Learning HCI Techniques: A Case Study on Interviews for Personas},
  author = {Barambones, Jose and Moral, Cristian and de Antonio, Angélica and Imbert, Ricardo and Mart{\'\i}nez-Normand, Lo{\"\i}c and Villalba-Mora, Elena},
  journal = {IEEE Transactions on Learning Technologies},
  volume = {17},
  pages = {1486--1501},
  year = {2024},
  doi = {10.1109/TLT.2024.3386095}
}

@inproceedings{budhiraja2024its,
  title = {“It’s not like Jarvis, but it’s pretty close!” - Examining ChatGPT’s Usage among Undergraduate Students in Computer Science},
  author = {Budhiraja, R and Joshi, I and Challa, J S and Akolekar, H D and Kumar, D},
  booktitle = {Proceedings of the 26th Australasian Computing Education Conference},
  pages = {124--133},
  year = {2024},
  publisher = {Association for Computing Machinery},
  doi = {10.1145/3636243.3636257}
}

@incollection{cope2024platformed,
  title = {Platformed Learning: Reshaping Education in the Era of Learning Management Systems},
  author = {Cope, Bill and Kalantzis, Mary},
  booktitle = {Critical EdTech Studies and Digital Platforms in Higher Education: Varieties of Platformisation},
  editor = {Thomas, Duncan A and Laterza, Vito},
  publisher = {Palgrave Macmillan},
  address = {London},
  year = {2024}
}

@article{ding2023students,
  title = {Students' perceptions of using ChatGPT in a physics class as a virtual tutor},
  author = {Ding, Lei and Li, Tao and Jiang, Shuai and et al.},
  journal = {International Journal of Educational Technology in Higher Education},
  volume = {20},
  number = {63},
  year = {2023},
  doi = {10.1186/s41239-023-00434-1},
  publisher = {Springer}
}

@inproceedings{frazier2024customizing,
  title = {Customizing ChatGPT to Help Computer Science Principles Students Learn Through Conversation},
  author = {Frazier, Matthew and Damevski, Kostadin and Pollock, Lori},
  booktitle = {Proceedings of the 2024 on Innovation and Technology in Computer Science Education V. 1 (ITiCSE 2024)},
  year = {2024},
  publisher = {Association for Computing Machinery},
  pages = {633--639},
  doi = {10.1145/3649217.3653570}
}

@article{garcia2025teaching,
  title = {Teaching and learning computer programming using ChatGPT: A rapid review of literature amid the rise of generative AI technologies},
  author = {Garcia, M B},
  journal = {Education and Information Technologies},
  volume = {30},
  pages = {16721--16745},
  year = {2025},
  doi = {10.1007/s10639-025-13452-5}
}

@article{haindl2024students,
  title = {Students’ Experiences of Using ChatGPT in an Undergraduate Programming Course},
  author = {Haindl, Philipp and Weinberger, Gerald},
  journal = {IEEE Access},
  volume = {12},
  pages = {43519--43529},
  year = {2024},
  doi = {10.1109/ACCESS.2024.3380909},
  publisher = {IEEE}
}

@inproceedings{joshi2024chatgpt,
  title = {ChatGPT in the Classroom: An Analysis of Its Strengths and Weaknesses for Solving Undergraduate Computer Science Questions},
  author = {Joshi, Ipsita and Budhiraja, Riya and Dev, Haimonti and Kadia, Jai and Ataullah, Mohammed Omar and Mitra, Snehil and Akolekar, Harshil D and Kumar, Divya},
  booktitle = {Proceedings of the 55th ACM Technical Symposium on Computer Science Education V. 1},
  pages = {625--631},
  year = {2024},
  publisher = {Association for Computing Machinery},
  doi = {10.1145/3626252.3630803}
}

@article{jost2024impact,
  title = {The impact of large language models on programming education and student learning outcomes},
  author = {Jošt, Gregor and Taneski, Vlatko and Karakatič, Sašo},
  journal = {Applied Sciences},
  volume = {14},
  number = {10},
  pages = {1--15},
  year = {2024},
  publisher = {MDPI},
  doi = {10.3390/app14104115}
}

@inproceedings{kazemitabaar2023studying,
  title = {Studying the effect of AI Code Generators on Supporting Novice Learners in Introductory Programming},
  author = {Kazemitabaar, Majeed and Chow, Justin and Ma, Carl Ka To and Ericson, Barbara J and Weintrop, David and Grossman, Tovi},
  booktitle = {Proceedings of the 2023 CHI Conference on Human Factors in Computing Systems (CHI '23)},
  year = {2023},
  publisher = {Association for Computing Machinery},
  pages = {1--23},
  doi = {10.1145/3544548.3580919}
}

@inproceedings{kazemitabaar2024how,
  title = {How Novices Use LLM-based Code Generators to Solve CS1 Coding Tasks in a Self-Paced Learning Environment},
  author = {Kazemitabaar, Majeed and Hou, Xinying and Henley, Austin and Ericson, Barbara Jane and Weintrop, David and Grossman, Tovi},
  booktitle = {Proceedings of the 23rd Koli Calling International Conference on Computing Education Research},
  year = {2024},
  publisher = {Association for Computing Machinery}
}

@article{kumar2024guiding,
  title = {Guiding Students in Using LLMs in Supported Learning Environments: Effects on Interaction Dynamics, Learner Performance, Confidence, and Trust},
  author = {Kumar, Harsh and Musabirov, Ilya and Reza, Mohi and Shi, Jiakai and Wang, Xinyuan and Williams, Joseph Jay and Kuzminykh, Anastasia and Liut, Michael},
  journal = {Proc. ACM Hum.-Comput. Interact. 8, CSCW2, Article 499},
  pages = {1--30},
  year = {2024},
  doi = {10.1145/3687038},
  publisher = {ACM}
}

@inproceedings{liang2024large,
  title = {A Large-Scale Survey on the Usability of AI Programming Assistants: Successes and Challenges},
  author = {Liang, Jenny T and Yang, Chenyang and Myers, Brad A},
  booktitle = {Proceedings of the IEEE/ACM 46th International Conference on Software Engineering (ICSE '24)},
  year = {2024},
  publisher = {Association for Computing Machinery}
}

@inproceedings{lyu2024evaluating,
  title = {Evaluating the Effectiveness of LLMs in Introductory Computer Science Education: A Semester-Long Field Study},
  author = {Lyu, Wenhan and Wang, Yimeng and Chung, Tingting (Rachel) and Sun, Yifan and Zhang, Yixuan},
  booktitle = {Proceedings of the Eleventh ACM Conference on Learning @ Scale (L@S '24)},
  year = {2024},
  publisher = {Association for Computing Machinery}
}

@book{misanchuk2023chatgpt,
  title = {ChatGPT in STEM Teaching: An introduction to using LLM-based tools in Higher Ed},
  author = {Misanchuk, Melanie},
  year = {2023},
  publisher = {eCampusOntario}
}

@inproceedings{nguyen2024how,
  title = {How Beginning Programmers and Code LLMs (Mis)read Each Other},
  author = {Nguyen, Sydney and Babe, Hannah McLean and Zi, Yangtian and Guha, Arjun and Anderson, Carolyn Jane and Feldman, Molly Q},
  booktitle = {Proceedings of the 2024 CHI Conference on Human Factors in Computing Systems (CHI '24)},
  year = {2024},
  publisher = {Association for Computing Machinery}
}

@inproceedings{oliveira2025can,
  title = {'Can You Refactor This for Me?': Investigating How Students Use ChatGPT in Code Refactoring Exercises},
  author = {Carneiro Oliveira, Eduardo and Keuning, Hieke and Jeuring, Johan},
  booktitle = {Proceedings of the 30th ACM Conference on Innovation and Technology in Computer Science Education V. 2 (ITiCSE 2025)},
  year = {2025},
  publisher = {Association for Computing Machinery}
}

@inproceedings{penney2023assessing,
  title={Assessing and Designing LLM-Based Conversational Agents for Introductory Computer Science Education},
  author={Penney, T. and Stephenson, B. and Denny, P. and Luxton-Reilly, A. and Petersen, A.},
  booktitle={Proceedings of the 18th Workshop in Primary and Secondary Computing Education (WiPSCE)},
  pages={1--11},
  year={2023},
  publisher={ACM},
  doi={10.1145/3621714.3621733}
}

@article{pirzado2024navigating,
  title = {Navigating the Pitfalls: Analyzing the Behavior of LLMs as a Coding Assistant for Computer Science Students—A Systematic Review of the Literature},
  author = {Pirzado, F A and Ahmed, A and Mendoza-Urdiales, R A and Terashima-Marin, H},
  journal = {IEEE Access},
  volume = {12},
  pages = {112605--112625},
  year = {2024},
  doi = {10.1109/ACCESS.2024.3443621}
}

@inproceedings{salminen2024using,
  title = {Using Cipherbot: An Exploratory Analysis of Student Interaction with an LLM-Based Educational Chatbot},
  author = {Salminen, Joni and Jung, Soon-gyo and Medina, Johanne and Aldous, Kholoud and Azem, Jinan and Akhtar, Waleed and Jansen, Bernard J},
  booktitle = {Proceedings of the Eleventh ACM Conference on Learning @ Scale (L@S '24)},
  year = {2024},
  publisher = {Association for Computing Machinery}
}

@article{shoufan2023can,
  title = {Can Students without Prior Knowledge Use ChatGPT to Answer Test Questions? An Empirical Study},
  author = {Shoufan, Abdulhadi},
  journal = {ACM Trans. Comput. Educ.},
  volume = {23},
  number = {4},
  year = {2023},
  publisher = {Association for Computing Machinery},
  doi = {10.1145/3628162}
}

@article{su2023collaborating,
  title = {Collaborating with ChatGPT in argumentative writing classrooms},
  author = {Su, Yu-Fang and Lin, Ya-Fen and Lai, Chia-Wen},
  journal = {Assessing Writing},
  volume = {57},
  pages = {100752},
  year = {2023},
  publisher = {Elsevier}
}

@article{wu2022time,
  title = {Time pressure changes how people explore and respond to uncertainty},
  author = {Wu, Caroline M and Schulz, Esther and Pleskac, Timothy J and Speekenbrink, Maarten},
  journal = {Scientific Reports},
  volume = {12},
  number = {1},
  pages = {4122},
  year = {2022},
  publisher = {Nature Publishing Group},
  doi = {10.1038/s41598-022-07901-1}
}

@inproceedings{vadaparty2024cs1,
  title = {CS1-LLM: Integrating LLMs into CS1 Instruction},
  author = {Vadaparty, Annapurna and Zingaro, Daniel and Smith IV, David H and Padala, Mounika and Alvarado, Christine and Benario, Jamie Gorson and Porter, Leo},
  booktitle = {Proceedings of the 2024 on Innovation and Technology in Computer Science Education V. 1 (ITiCSE 2024)},
  year = {2024},
  publisher = {Association for Computing Machinery}
}

@article{zhou2024developing,
  title = {Developing an interaction framework for human-large language models collaboration in creative tasks: Insights from UX professionals’ communication with ChatGPT},
  author = {Zhou, Zihan and Gao, Wei and Li, Yifei and Yu, Jie},
  journal = {Available at SSRN},
  volume = {6},
  pages = {1--47},
  year = {2024},
  doi = {10.2139/ssrn.4853257}
}

@book{bloom1964taxonomy,
  title={Taxonomy of educational objectives},
  author={Bloom, Benjamin Samuel and Engelhart, Max D and Furst, Edward J and Hill, Walker H and Krathwohl, David R},
  volume={2},
  year={1964},
  publisher={Longmans, Green New York}
}

@article{sellen2025effects,
author = {Sellen, Abigail},
title = {Effects of LLM Use and Note-Taking On Reading Comprehension and Memory: A Randomised Experiment in Secondary Schools},
year = {2025},
month = {February},
journal = {SSRN Electronic Journal},
}

@incollection{SWELLER201137,
title = {CHAPTER TWO - Cognitive Load Theory},
editor = {Jose P. Mestre and Brian H. Ross},
series = {Psychology of Learning and Motivation},
publisher = {Academic Press},
volume = {55},
pages = {37-76},
year = {2011},
issn = {0079-7421},
doi = {https://doi.org/10.1016/B978-0-12-387691-1.00002-8},
url = {https://www.sciencedirect.com/science/article/pii/B9780123876911000028},
author = {John Sweller},
}

@article{ordonez1997decisions,
  title={Decisions under time pressure: How time constraint affects risky decision making},
  author={Ordonez, Lisa and Benson III, Lehman},
  journal={Organizational Behavior and Human Decision Processes},
  volume={71},
  number={2},
  pages={121--140},
  year={1997},
  publisher={Elsevier}
}

@article{cabrera2023improving,
  title = {Improving Human-AI Collaboration With Descriptions of AI Behavior},
  author = {Cabrera, {\'A}ngel Alexander and Perer, Adam and Hong, Jason I},
  journal = {Proceedings of the ACM on Human-Computer Interaction},
  volume = {7},
  number = {CSCW1},
  pages = {136},
  year = {2023},
  publisher = {ACM},
  doi = {10.1145/3579612}
}

@article{kruskal1952use,
  title = {Use of ranks in one-criterion variance analysis},
  author = {Kruskal, William H and Wallis, W Allen},
  journal = {Journal of the American Statistical Association},
  volume = {47},
  pages = {583--621},
  year = {1952},
  publisher = {Taylor & Francis}
}

@article{mann1947test,
  title = {On a test of whether one of two random variables is stochastically larger than the other},
  author = {Mann, Henry B and Whitney, Donald R},
  journal = {Annals of Mathematical Statistics},
  volume = {18},
  pages = {50--60},
  year = {1947},
  publisher = {Institute of Mathematical Statistics}
}

@article{holm1979simple,
  title = {A simple sequentially rejective multiple test procedure},
  author = {Holm, Sture},
  journal = {Scandinavian Journal of Statistics},
  volume = {6},
  number = {2},
  pages = {65--70},
  year = {1979},
  publisher = {Wiley}
}

@book{kommers1992cognitive,
  title = {Cognitive Tools for Learning},
  editor = {Kommers, Piet A. M. and Jonassen, David H. and Mayes, J. Terry and Ferreira, Alcindo},
  year = {1992},
  publisher = {Springer Berlin, Heidelberg},
  series = {NATO ASI Subseries F},
  doi = {10.1007/978-3-642-77222-1}
}

@inproceedings{ma2025towards,
  title = {Towards Human-AI Deliberation: Design and Evaluation of LLM-Empowered Deliberative AI for AI-Assisted Decision-Making},
  author = {Ma, Shuai and Chen, Qiaoyi and Wang, Xinru and Zheng, Chengbo and Peng, Zhenhui and Yin, Ming and Ma, Xiaojuan},
  booktitle = {Proceedings of the 2025 CHI Conference on Human Factors in Computing Systems},
  year = {2025},
  publisher = {Association for Computing Machinery}
}

@article{reinhold2024learning,
  title = {Learning Mechanisms Explaining Learning With Digital Tools in Educational Settings: a Cognitive Process Framework},
  author = {Reinhold, Felicitas and Leuders, Timo and Loibl, Katharina and et al.},
  journal = {Educational Psychology Review},
  volume = {36},
  number = {14},
  year = {2024},
  doi = {10.1007/s10648-024-09845-6}
}

@article{spitzer2025human,
  title = {Human Delegation Behavior in Human-AI Collaboration: The Effect of Contextual Information},
  author = {Spitzer, Philipp and Holstein, Joshua and Hemmer, Patrick and V{\"o}ssing, Michael and K{\"u}hl, Niklas and Martin, Dominik and Satzger, Gerhard},
  journal = {Proceedings of the ACM on Human-Computer Interaction},
  volume = {9},
  number = {2},
  pages = {101},
  year = {2025},
  publisher = {ACM},
  doi = {10.1145/3710999}
}

@inproceedings{tankelevitch2024metacognitive,
  title = {The Metacognitive Demands and Opportunities of Generative AI},
  author = {Tankelevitch, Lev and Kewenig, Viktor and Simkute, Auste and Scott, Ava Elizabeth and Sarkar, Advait and Sellen, Abigail and Rintel, Sean},
  booktitle = {Proceedings of the 2024 CHI Conference on Human Factors in Computing Systems},
  year = {2024},
  publisher = {Association for Computing Machinery}
}

@article{van2010cognitive,
  title = {Cognitive Load Theory: Advances in Research on Worked Examples, Animations, and Cognitive Load Measurement},
  author = {van Gog, Ton and Paas, Fred and Sweller, John},
  journal = {Educational Psychology Review},
  volume = {22},
  number = {3},
  pages = {375--378},
  year = {2010},
  publisher = {Springer},
  doi = {10.1007/s10648-010-9145-4}
}

@book{vygotsky1978mind,
  title = {Mind in Society: Development of Higher Psychological Processes},
  author = {Vygotsky, Lev Semenovich},
  editor = {Cole, Michael and John-Steiner, Vera and Scribner, Sylvia and Souberman, Ellen},
  year = {1978},
  publisher = {Harvard University Press},
  address = {Cambridge, MA},
  doi = {10.2307/j.ctvjf9vz4}
}

\newpage

\appendix
\section*{\textbf{\LARGE APPENDIX}}
\section*{\textbf{Table of Contents}}
\noindent A GPT Annotation Protocol \dotfill 23

\noindent B Assignment Grading Rubric \dotfill 25

\newpage

\section{GPT Annotation Protocols}
\label{app: gpt_training_prompt}
\subsection{Prompt Annotation Instructions}
The following instructions were used to guide coders (both human and LLM-assisted) in annotating student prompts. They detail the procedure, codebook, and output format required for consistent coding.

\textbf{Task Overview.}  
You are an expert qualitative coder. You will analyze student chat transcripts (with AI) using the provided codebook and produce outputs that can be pasted \textbf{directly} into the Excel template provided.

\textbf{File / Template.}  
\begin{itemize}
    \item Template: \texttt{prompt-annotation-template.xlsx}, Sheet1
    \item Required columns: 
    \begin{enumerate}
        \item Student ID
        \item Round ID
        \item Prompt Text
        \item Theme 5
        \item Theme 6
        \item Theme 7
        \item Theme 8
        \item Theme 9
        \item Theme 10
        \item Theme 11
        \item Theme 12
        \item Notes
    \end{enumerate}
\end{itemize}

\textbf{Codebook Definitions.}  
\begin{itemize}
    \item \textbf{Theme 5. Full code solution:} Prompt explicitly or implicitly requested AI to write the entire code solution.  
    Example: \textit{“Make a program using the pdf document instructions and the outline provided in the .py file.”}

    \item \textbf{Theme 6. Step/feature code:} Prompt requested AI to write code for a certain step or feature.  
    Example: \textit{“Please give me a Gameplay Loop in Python that fulfills these requirements…”}

    \item \textbf{Theme 7. Error/missing pieces:} Prompt requested code fix or modification by describing an error or pointing out missing pieces.  
    Example: \textit{“there’s only one function.”}

    \item \textbf{Theme 8. General fix/modification:} Prompt requested modification without specific requirements.  
    Example: code dump without explanation.

    \item \textbf{Theme 9. Explanation:} Prompt requested AI to explain generated code.  
    Example: \textit{“Can you tell me for which part is which code…”}

    \item \textbf{Theme 10. Synthesis:} Prompt requested AI to synthesize outcomes from earlier conversations.  
    Example: \textit{“update the high score part.”}

    \item \textbf{Theme 11. Other Guidance:} Prompt did not ask for an answer or implementation.  
    Example: \textit{“awesome! can I directly paste this after the code you gave me before?”}

    \item \textbf{Theme 12. Response to AI:} Prompt answered or confirmed an AI suggestion.  
    Example: \textit{“yes, it is look at it.”}
\end{itemize}

\textbf{Procedure.}  
\begin{enumerate}
    \item Each student message = one row.
    \item Fill in Student ID and sequential Round ID (R1, R2, …).
    \item Copy the prompt verbatim under “Prompt Text.”
    \item Mark Theme 5–12 as 1 if applicable, 0 otherwise (multiple 1s allowed).
    \item Use Notes column only for optional clarifications.
\end{enumerate}

\textbf{Output Format (CSV).}  
\begin{verbatim}
"Student ID","Round ID","Prompt Text","Theme 5","Theme 6","Theme 7",
"Theme 8","Theme 9","Theme 10","Theme 11","Theme 12","Notes"
11,"R1","Here is my assignment instruction. Can you add the replay feature?",
0,1,0,0,0,0,0,0,""
11,"R2","Yes please add that",
0,0,0,0,0,0,0,1,"Acknowledgement to AI"
\end{verbatim}

\section{Grading Rubrics}
\label{app:grading_rubics}
\begin{table}[h!]
\centering
\footnotesize % Use a smaller font size for this large table

% The @{} removes extra space at the table's edges for better alignment.
\begin{tabular}{@{} p{0.12\textwidth} p{0.22\textwidth} p{0.22\textwidth} p{0.22\textwidth} p{0.22\textwidth} @{}}
\toprule
\textbf{Criteria} & \textbf{Poor (0 pts)} & \textbf{Fair (3 pts)} & \textbf{Good (7 pts)} & \textbf{Excellent (10 pts)} \\
\midrule
\multicolumn{5}{l}{\textbf{Gameplay Loop}} \\ \cmidrule(r){1-5}
Guessing Mechanism \& Loop & Guessing loop is missing or does not function. Player cannot input guesses. & Guessing loop is present but has significant errors (e.g., infinite loop, incorrect termination). Player input is buggy. & Guessing loop functions, allows player input, and generally terminates, but may have minor logic flaws. & Guessing loop is correctly implemented, allows player input, and terminates accurately based on win/loss conditions. \\ \addlinespace
Feedback (High/Low/Correct) & No feedback provided, or feedback is consistently incorrect. & Feedback is provided but is often incorrect or missing for some conditions (e.g., only ``too high'' works). & Feedback for ``too high,'' ``too low,'' and correct guesses is mostly accurate. & Feedback (``Too high!'', ``Too low!'', congratulatory message) is consistently accurate and clearly presented. \\ \addlinespace
Attempt Tracking \& Limits & Attempts are not tracked, or the limit is not enforced. & Attempts are tracked incorrectly, or the game does not end when attempts run out. & Attempts are tracked, and the game ends when attempts run out, but there might be off-by-one errors in counting. & Attempts are accurately tracked and displayed (if required by output). Game correctly ends when attempts are exhausted. \\ \addlinespace
Input Validation (Guess) & Program crashes if a non-numeric guess is entered. & Handles non-numeric guesses poorly (e.g., error message but loop breaks, or incorrect behavior). & Handles non-numeric guesses with an error message but might not re-prompt correctly or deduct an attempt unfairly. & Gracefully handles non-numeric guess input, informs the user, and allows them to try again without penalizing attempts unfairly. \\
\midrule
\multicolumn{5}{l}{\textbf{Game Logic}} \\ \cmidrule(r){1-5}
Win Condition \& Attempt Count & Does not correctly identify a win, or attempts taken for a win are not recorded. & Identifies a win but records the number of attempts taken incorrectly (e.g., always 0, or max attempts). & Correctly identifies a win and records the number of attempts taken, with minor inaccuracies in some edge cases. & Correctly identifies a win and accurately records the number of attempts taken for the successful round. \\ \addlinespace
Loss Condition & Does not correctly identify when the player runs out of attempts. & Identifies a loss (out of attempts) but may still allow further guesses or behaves erratically. & Correctly identifies a loss when attempts are exhausted. & Correctly identifies a loss, stops the guessing, and proceeds to the next step (e.g., record result, play again). \\ \addlinespace
Recording Results per Round & Player name and attempts are not stored after each round. & Player name and/or attempts are stored, but data is incorrect, incomplete, or lost between rounds. & Player name and attempts are stored for each round in the result list, mostly correctly. & Player name and attempts for each round are accurately stored in the result list as per specifications. \\ \addlinespace
Game State Reset for New Round & Game state (target number, attempts) is not reset for a new round. & Game state is partially reset, leading to issues in subsequent rounds (e.g., same target number, incorrect attempts). & Game state is mostly reset correctly for a new round. & Game state (new random number, full attempts for the chosen difficulty) is correctly and fully reset for each new round. \\
\midrule
\multicolumn{5}{l}{\textbf{Play Again \& Game End}} \\ \cmidrule(r){1-5}
``Play Again?'' Prompt \& Logic & Does not ask the player if they want to play again, or the game always ends/restarts. & Asks to play again, but input handling is flawed (e.g., only 'yes' works, 'no' restarts, or invalid input breaks). & Asks to play again and handles 'yes'/'no' correctly but may not handle other inputs gracefully. & Asks to play again, correctly processes 'yes' (starts new round) and 'no' (ends game session), and handles invalid input. \\ \addlinespace
Displaying Final Results Summary & No summary of results is displayed at the end of the game session. & A summary is attempted but is missing key information (names/attempts), is formatted poorly, or shows incorrect data. & Displays a summary of results (player name and attempts for each round) from the result list, with minor formatting issues. & Clearly and accurately displays the summary of all rounds (player name and attempt count for each) as per the example output. \\
\midrule
\multicolumn{5}{l}{\textbf{Coding Style \& Readability}} \\ \cmidrule(r){1-5}
Adherence to Skeleton \& Comments & Code significantly deviates from the skeleton. Implemented sections are uncommented or comments are unhelpful. & Follows skeleton loosely. Minimal comments, or comments do not clarify the logic of the implemented sections. & Generally follows the provided code skeleton. Implemented sections have adequate comments explaining the basic logic. & Closely follows the provided code skeleton. Implemented sections are well-commented, clearly explaining the purpose and logic of the code. \\ \addlinespace
Variable Naming \& Indentation & Variable names are unclear (e.g., x, y). Indentation is inconsistent or incorrect, making code hard to read. & Some variable names are unclear. Indentation has inconsistencies. & Variable names are mostly descriptive. Indentation is generally correct. Code is readable. & Variable names are clear and descriptive. Code is consistently and correctly indented, adhering to Python conventions. \\ \addlinespace
Code Clarity \& Structure & Logic is convoluted, difficult to follow. Unnecessary complexity or redundancy. & Logic is somewhat difficult to follow in places. Some redundant code or inefficient structures. & Logic is generally clear. Code structure is reasonable. Minor areas could be more concise or clear. & Logic is clear, concise, and easy to understand. Code is well-structured within the start\_game function and any helper logic. \\
\bottomrule
\end{tabular}
\caption{Rubric used for grading code submissions}
\label{tab:code-rubric}
\end{table}

\end{document}